\documentclass[review]{elsarticle}

\usepackage{lineno,hyperref,subfigure,color}
\modulolinenumbers[1]
\journal{Astroparticle Physics}
\begin{document}
\begin{frontmatter}

\title{The Energy Spectrum of Cosmic Rays above 10$^{17.2}$ eV Measured by the Fluorescence Detectors of the Telescope Array Experiment in Seven Years}

\author[utah]{R.U.~Abbasi}
\address[utah]{High Energy Astrophysics Institute and Department of
Physics and Astronomy, University of Utah, Salt Lake City, Utah, USA}

\author[saitama]{M.~Abe}
\address[saitama]{The Graduate School of Science and Engineering, Saitama
University, Saitama, Saitama, Japan}

\author[utah]{T.~Abu-Zayyad}  

\author[utah]{M.~Allen} 

\author[titec]{R.~Azuma} 
\address[titec]{Graduate School of Science and Engineering, Tokyo
Institute of Technology, Meguro, Tokyo, Japan}

\author[utah]{E.~Barcikowski} 

\author[utah]{J.W.~Belz}

\author[utah]{D.R.~Bergman} 

\author[utah]{S.A.~Blake} 

\author[utah]{R.~Cady} 


\author[hanyang]{B.G.~Cheon}
\address[hanyang]{Department of Physics and The Research Institute of
Natural Science, Hanyang University, Seongdong-gu, Seoul, Korea}

\author[tus]{J.~Chiba}
\address[tus]{Department of Physics, Tokyo University of Science, Noda,
Chiba, Japan} 

\author[kinki]{M.~Chikawa}
\address[kinki]{Department of Physics, Kinki University, Higashi Osaka,
Osaka, Japan}

\author[yonsei]{W.R.~Cho} 
\address[yonsei]{Department of Physics, Yonsei University, Seodaemun-gu,
Seoul, Korea}

\author[icrr]{T.~Fujii\fnref{fujii}\corref{mycorrespondingauthor}}
\address[icrr]{Institute for Cosmic Ray Research, University of Tokyo,
Kashiwa, Chiba, Japan}
\cortext[mycorrespondingauthor]{Corresponding Author}
\fntext[fujii]{Now at University of Chicago, USA}
\ead{fujii@icrr.u-tokyo.ac.jp}

\author[icrr,ipmu]{M.~Fukushima}
\address[ipmu]{Kavli Institute for the Physics and Mathematics of the
Universe (WPI), Todai Institutes for Advanced Study, the University of
Tokyo, Kashiwa, Chiba, Japan}

\author[ocu]{T.~Goto}
\address[ocu]{Graduate School of Science, Osaka City University, Osaka,
Osaka, Japan}

\author[utah]{W.~Hanlon} 

\author[ocu]{Y.~Hayashi} 

\author[kanagawa]{N.~Hayashida} 
\address[kanagawa]{Faculty of Engineering, Kanagawa University, Yokohama,
Kanagawa, Japan}

\author[kanagawa]{K.~Hibino} 

\author[yamanashi]{K.~Honda}
\address[yamanashi]{Interdisciplinary Graduate School of Medicine and
Engineering, University of Yamanashi, Kofu, Yamanashi, Japan}

\author[icrr]{D.~Ikeda} 

\author[saitama]{N.~Inoue} 

\author[yamanashi]{T.~Ishii} 

\author[titec]{R.~Ishimori}

\author[riken]{H.~Ito}
\address[riken]{Astrophysical Big Bang Laboratory, RIKEN, Wako, Saitama,
Japan}

\author[utah]{D.~Ivanov} 

\author[utah]{C.C.H.~Jui} 

\author[tcu]{K.~Kadota} 
\address[tcu]{Department of Physics, Tokyo City University,
Setagaya-ku, Tokyo, Japan}

\author[titec]{F.~Kakimoto}

\author[inr]{O.~Kalashev} 
\address[inr]{Institute for Nuclear Research of the Russian Academy of
Sciences, Moscow, Russia}

\author[waseda]{K.~Kasahara} 
\address[waseda]{Advanced Research Institute for Science and Engineering,
Waseda University, Shinjuku-ku, Tokyo, Japan}

\author[chiba]{H.~Kawai} 
\address[chiba]{Department of Physics, Chiba University, Chiba, Chiba,
Japan}

\author[ocu]{S.~Kawakami} 

\author[saitama]{S.~Kawana} 

\author[icrr]{K.~Kawata} 

\author[icrr]{E.~Kido} 

\author[hanyang]{H.B.~Kim}

\author[utah]{J.H.~Kim} 

\author[ulsan]{J.H.~Kim} 
\address[ulsan]{Department of Physics, School of Natural Sciences, Ulsan
National Institute of Science and Technology, UNIST-gil, Ulsan, Korea}

\author[titec]{S.~Kitamura}

\author[titec]{Y.~Kitamura}

\author[inr]{V.~Kuzmin\fnref{kuzmin}}
\fntext[kuzmin]{Deceased}

\author[yonsei]{Y.J.~Kwon} 

\author[utah]{J.~Lan}
\address[]{High Energy Astrophysics Institute and Department of
Physics and Astronomy, University of Utah, Salt Lake City, Utah, USA}


\author[utah]{J.P.~Lundquist} 

\author[yamanashi]{K.~Machida} 

\author[ipmu]{K.~Martens} 

\author[kek]{T.~Matsuda} 
\address[kek]{Institute of Particle and Nuclear Studies, KEK, Tsukuba,
Ibaraki, Japan}

\author[ocu]{T.~Matsuyama} 

\author[utah]{J.N.~Matthews} 

\author[ocu]{M.~Minamino} 

\author[yamanashi]{Y.~Mukai} 

\author[utah]{I.~Myers}

\author[saitama]{K.~Nagasawa}

\author[riken]{S.~Nagataki}

\author[kochi]{T.~Nakamura} 
\address[kochi]{Faculty of Science, Kochi University, Kochi, Kochi,
Japan}

\author[icrr]{T.~Nonaka} 

\author[kinki]{A.~Nozato} 

\author[ocu]{S.~Ogio} 

\author[titec]{J.~Ogura}

\author[icrr]{M.~Ohnishi} 
\author[icrr]{H.~Ohoka} 
\author[icrr]{K.~Oki} 

\author[rits]{T.~Okuda} 
\address[rits]{Department of Physical Sciences, Ritsumeikan University,
Kusatsu, Shiga, Japan}

\author[fukuoka]{M.~Ono} 
\address[fukuoka]{Department of Physics, Kyushu University, Fukuoka,
Fukuoka, Japan}

\author[chubu]{A.~Oshima} 
\address[chubu]{Engineering Science Laboratory, Chubu University,
Kasugai, Aichi, Japan}

\author[waseda]{S.~Ozawa} 

\author[sungkyunkwan]{I.H.~Park} 
\address[sungkyunkwan]{Department of Physics, Sungkyunkwan University,
Jang-an-gu, Suwon, Korea}

\author[inr,sai]{M.S.~Pshirkov} 
\address[sai]{Sternberg Astronomical Institute,  Moscow M.V. Lomonosov
State University, Moscow, Russia}

\author[utah]{D.C.~Rodriguez} 

\author[inr]{G.~Rubtsov} 

\author[ulsan]{D.~Ryu} 

\author[icrr]{H.~Sagawa} 

\author[ocu]{N.~Sakurai} 

\author[rutgers]{L.M.~Scott}
\address[rutgers]{Department of Physics and Astronomy, Rutgers University
-- The State University of New Jersey, Piscataway, New Jersey, USA}

\author[utah]{P.D.~Shah} 

\author[yamanashi]{F.~Shibata} 

\author[icrr]{T.~Shibata} 

\author[icrr]{H.~Shimodaira} 

\author[hanyang]{B.K.~Shin}

\author[icrr]{H.S.~Shin}

\author[utah]{J.D.~Smith} 

\author[utah]{P.~Sokolsky}

\author[utah]{R.W.~Springer}  

\author[utah]{B.T.~Stokes} 

\author[utah,rutgers]{S.R.~Stratton} 

\author[utah]{T.A.~Stroman} 

\author[saitama]{T.~Suzawa}

\author[tus]{M.~Takamura}

\author[icrr]{M.~Takeda} 

\author[icrr]{R.~Takeishi}

\author[eri]{A.~Taketa} 
\address[eri]{Earthquake Research Institute, University of Tokyo,
Bunkyo-ku, Tokyo, Japan}

\author[icrr]{M.~Takita}

\author[kanagawa]{Y.~Tameda} 

\author[ocu]{H.~Tanaka} 

\author[hiroshima]{K.~Tanaka} 
\address[hiroshima]{Graduate School of Information Sciences, Hiroshima City
University, Hiroshima, Hiroshima, Japan}

\author[kek]{M.~Tanaka} 

\author[utah]{S.B.~Thomas} 

\author[utah]{G.B.~Thomson} 

\author[brus,inr]{P.~Tinyakov} 
\address[brus]{Service de Physique Th\'eorique, Universit\'e Libre de Bruxelles, Brussels, Belgium}

\author[inr]{I.~Tkachev} 

\author[titec]{H.~Tokuno}

\author[shinshu]{T.~Tomida} 
\address[shinshu]{Department of Computer Science and Engineering, Shinshu
University, Nagano, Nagano, Japan}

\author[inr]{S.~Troitsky} 

\author[ocu]{Y.~Tsunesada}  

\author[titec]{K.~Tsutsumi}

\author[nirs]{Y.~Uchihori} 
\address[nirs]{National Institute of Radiological Science, Chiba, Chiba,
Japan}

\author[kanagawa]{S.~Udo}

\author[brus]{F.~Urban}

\author[utah]{G.~Vasiloff} 

\author[utah]{T.~Wong} 

\author[ocu]{R.~Yamane}

\author[kek]{H.~Yamaoka}

\author[eri]{K.~Yamazaki} 

\author[ewha]{J.~Yang}
\address[ewha]{Department of Physics, Ewha Womans University, 
Seodaaemun-gu, Seoul, Korea} 

\author[tus]{K.~Yashiro}

\author[ocu]{Y.~Yoneda} 

\author[chiba]{S.~Yoshida} 

\author[ehime]{H.~Yoshii} 
\address[ehime]{Department of Physics, Ehime University, Matsuyama,
Ehime, Japan}

\author[utah]{R.~Zollinger} 

\author[utah]{Z.~Zundel}

\begin{abstract} 
The Telescope Array (TA) experiment is the largest detector to observe ultra-high-energy cosmic rays in the northern hemisphere.
The fluorescence detectors at two stations of TA are newly constructed and have now completed seven years of steady operation.
One advantage of monocular analysis of the fluorescence detectors is a lower energy threshold for cosmic rays than that of other techniques like stereoscopic observations or coincidences with the surface detector array, allowing the measurement of an energy spectrum covering three orders of magnitude in energy.
Analyzing data collected during those seven years, we report the energy spectrum of cosmic rays covering a broad range of energies above $10^{17.2}$~eV measured by the fluorescence detectors and a comparison with previously published results.
\end{abstract}


\end{frontmatter}


\section{Introduction}
\label{sec:introduction}
The energy spectrum of cosmic rays has been measured at energies from $10^{8}$~eV to beyond $10^{20}$~eV in the century elapsed since the discovery of cosmic rays using a balloon measurement~\cite{bib:history_cosmicray}.
In the energy range above 10$^{11}$~eV, the cosmic-ray flux is not affected by solar activity; the spectrum follows the power-law structure of $E^{-\gamma}$.
The spectral index, $\gamma$, is $\sim 2.7$ up to $10^{15.5}$~eV, where it changes to $\gamma \sim 3.0$. The spectrum softens slightly to $\gamma \sim 3.3$ around $10^{17}$~eV, and hardens again to $\gamma \sim 2.7$ just below $10^{19}$~eV. These three spectral-index discontinuities are called, in order of increasing energy, the ``knee,'' the ``second knee,'' and the ``ankle.''
Cosmic rays above 10$^{18}$~eV are called ultra-high-energy cosmic rays (UHECRs).
If we assume that UHECRs are proton-dominated, we expect UHE protons above 10$^{19.7}$~eV to interact with cosmic microwave background radiation (CMBR) via pion production, and the mean free path of UHE protons will be significantly reduced. Heavier nuclei also interact with the CMBR via photo-disintegration processes.  As a result, the energy spectrum above $10^{19.7}$~eV will be suppressed---the so-called GZK cutoff~\cite{bib:gzk1,bib:gzk2}.

Since UHECRs are the most energetic particles in the universe, their origins are ostensibly related to extremely energetic astronomical phenomena or other exotic processes, such as the decay or annihilation of super-heavy relic particles created in an early phase of the development of the universe. 
A transition of cosmic-ray origins from galactic to extragalactic sources might be happening around 10$^{17}$ eV.
Therefore, precise measurements of the energy spectrum covering a broad energy range and its structure are of utmost importance to understand the origin and propagation of cosmic rays.
However, at the highest energies ($\sim 10^{20}$~eV), the cosmic-ray flux becomes quite low, only one particle per century per km$^2$ area. A huge effective area and a large exposure are essential to measure cosmic rays with such energies.

\section{Telescope Array experiment}
\label{sec:ta}
The Telescope Array (TA) experiment is the largest cosmic-ray detector in the northern hemisphere~\cite{bib:TA}.
TA consists of 507 surface detectors (SDs) deployed on a square grid with 1.2-km spacing, covering an effective area of about 700~km$^2$~\cite{bib:TA_SD}; the SD array is
overlooked by 38 fluorescence detectors (FDs) at three locations~\cite{bib:htokuno_telescope}. 
One FD station, called ``Middle Drum'' (MD) and located northwest of the SD array, consists of 14 FDs previously used in the High Resolution Fly's Eye (HiRes) experiment~\cite{bib:hires}. 
Two other stations at the array's southeast and southwest, respectively called ``Black Rock Mesa'' (BRM) and ``Long Ridge'' (LR), each consist of 12 newly designed FDs \cite{bib:htokuno_telescope}. 

Because the duty cycle of the SDs is approximately 100\%, they have the greatest exposure, and hence the largest number of events, of any TA analysis. The SD measurement of the cosmic-ray energy spectrum above 10$^{18.2}$~eV, including a suppression above 10$^{19.7}$~eV consistent with the GZK cutoff, has been previously reported~\cite{bib:tasd_spectrum}. 
The energy spectrum observed by the MD station and a comparison with the HiRes experiment were reported after the first three years of operation~\cite{bib:md_spectrum}.
The energy spectrum above 10$^{18}$~eV analyzing data collected during three-and-a-half years of operation by the FDs was previously reported~\cite{bib:tafd_spectrum}.
In this paper, we report an update on the energy spectrum with double the statistics from seven years of observation, and extend the range of energies down to 10$^{17.2}$~eV.

The field of view (FOV) of the BRM and LR stations' FDs is approximately $18^\circ$ in azimuth and $15.5^\circ$ in elevation. At each station, 12 FDs are arranged in two layers of 6 telescopes each. The upper-layer telescopes are oriented to view the sky at elevations from $3^\circ$ to $18.5^\circ$, and the lower layer observes higher elevations, from  $17.5^\circ$ to $33^\circ$, for a combined elevation coverage of $3^\circ \sim 33^\circ$.
The directions of the telescopes are fanned out covering a combined $108^\circ$ in azimuth at each station, with BRM looking generally west and northwest, and LR looking east and northeast.
The BRM and LR FDs have spherical mirrors with a diameter of 3.3 m, which are composed of 18 hexagonal segments.
The camera design of these FD consists of 256 hexagonal photo-multiplier tubes (PMTs) arranged in a 16$\times$16 honeycomb array, installed at the focal plane of the spherical mirror.
The anode voltage of each PMT is digitized with a 12-bit, 40-MHz flash analog-to-digital converter (FADC). 
Sets of four adjacent time bins are summed to obtain an equivalent sampling rate of 10 MHz with a 14-bit dynamic range.
The trigger electronics can select a track pattern of triggered PMTs in real time to reduce accidental noise~\cite{bib:ytameda_trigger}.

The absolute gain of a few standard PMTs (2 or 3 installed in each camera) is measured by CRAYS (Calibration using RAYleigh Scattering) in the laboratory~\cite{bib:CRAYS}.
The gain of the other PMTs in the same camera can be monitored relative to the standard PMTs by comparing the flash intensities of xenon lamps which are installed in the center of each mirror~\cite{bib:htokuno_ta_calib,bib:yap}.
In order to calibrate the atmospheric conditions and relative gains among the three FD stations, a Central Laser Facility (CLF) is installed at the center of the TA site, a distance of 21~km from each station. An ultraviolet laser with a wavelength of 355~nm fires a 30-second burst of 300 vertical shots every 30~minutes~\cite{bib:ta_clf}. The laser energy is $\sim$4.0~mJ; the light scattered from the beam is equivalent to that produced by an equally distant UHECR with energy of $\sim 10^{19.5}$~eV.
The atmospheric transmittance was monitored at the start and end of daily operation of the FD by a Light Detection And Ranging (LIDAR) system by May~2012, which measured the back-scattered photons from an ultraviolet laser~\cite{bib:lidar_tomida}.
To categorize the extent of cloud cover, we use a weather code visually recorded by the operator at the northern station. These records are consistent with a picture taken by the infrared camera at the southeast station, and charge-coupled device (CCD) fisheye cameras installed at three locations around the experiment~\cite{bib:cloud}.

\section{Analysis procedure in monocular mode}
\begin{figure}[h]
  \centering
  \subfigure{\includegraphics[width=1.0\linewidth]{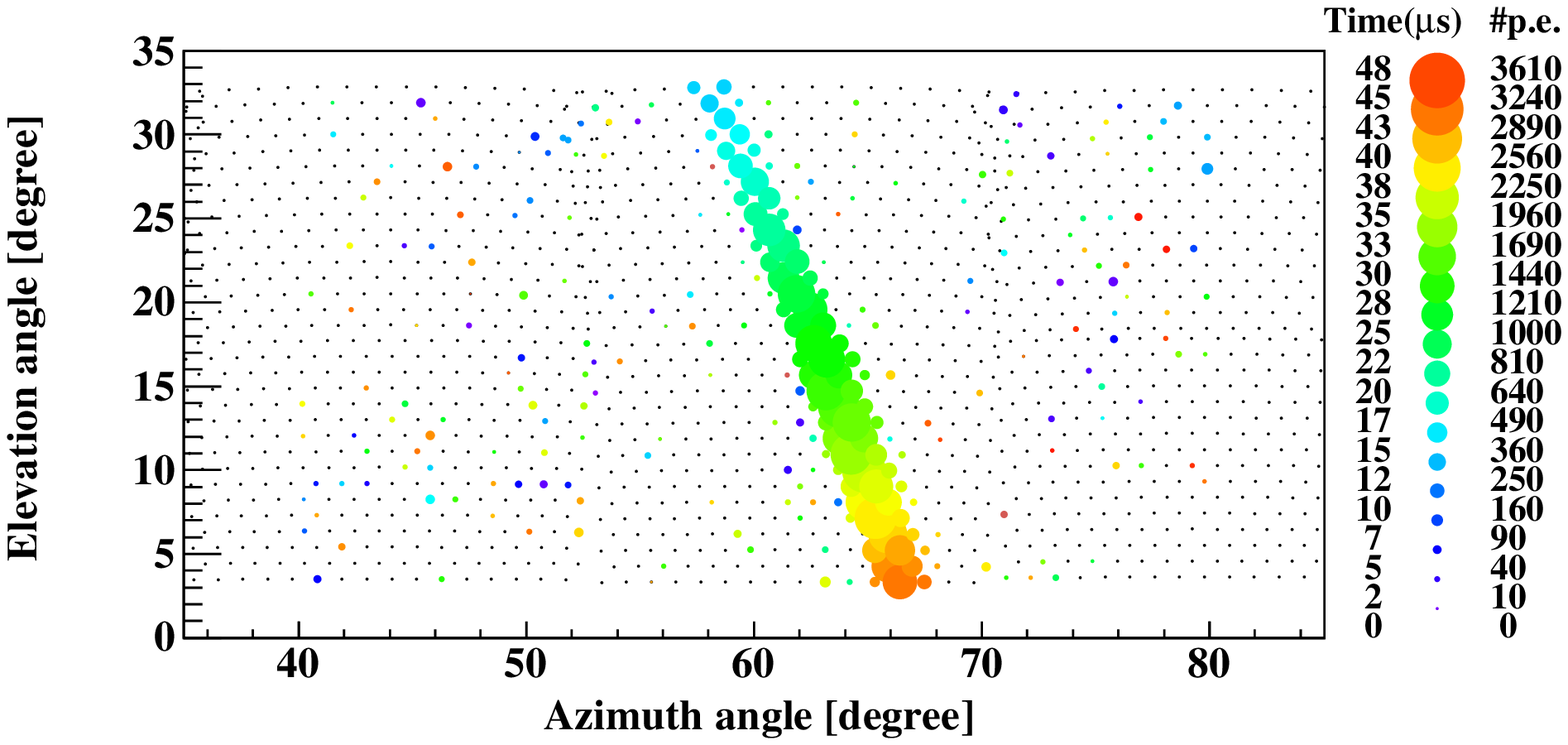}}
  \subfigure{\includegraphics[width=1.0\linewidth]{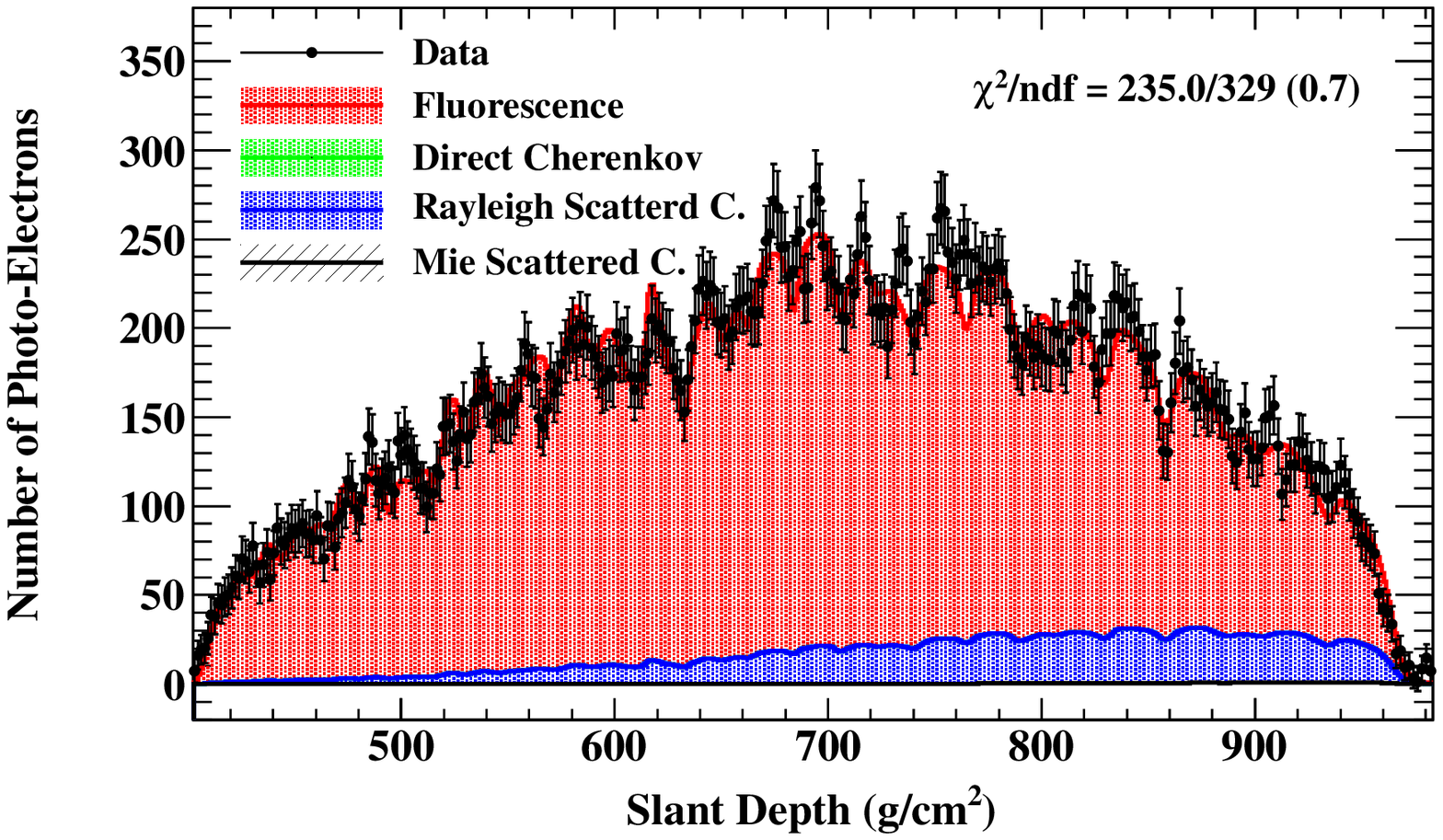}}
  \caption{UHECR event observed by the fluorescence detector. The top figure shows the PMT pointing directions and the brightness of signal (point size) and timing (point color). The bottom figure shows a sum of selected PMT waveforms as a function of slant depth (black plot), compared with the reconstructed result by the inverse Monte Carlo reconstruction (histograms). The inverse Monte Carlo method can reproduce the obtained signal at the camera, including the periodicity introduced by photo-cathode non-uniformity.}
  \label{fig:example}
\end{figure}

We analyze data collected at the BRM and LR stations using a monocular analysis, which is an analysis mode to reconstruct an extensive air shower (EAS) to obtain the primary particle's properties using the measured shower image by one FD station. The analysis procedure, which we summarize here, was described in detail in~\cite{bib:TA_IMC}.
First, PMTs having a signal from fluorescence light emitted at the EAS are identified by both an elevated signal-to-noise ratio and the proximity in space and time of several such PMTs.
To reconstruct the geometry of an observed EAS, the arrival time $t_i$ of the signal in each selected PMT $i$ is fitted by
\begin{equation}
t_i = t_{\rm{core}} + \frac{r_0}{c}\frac{\sin \Psi - \sin \alpha_i}{\sin (\Psi + \alpha_i)},
\end{equation}
where $\alpha_i$ is the angle formed by the $i^{\rm th}$ PMT's viewing direction and a direction vector from the FD station to the shower core (the shower axis's impact point on the ground), $\Psi$ is the angle on the shower-detector plane formed by the shower axis and the direction to the shower core, $t_{\rm{core}}$ is the time when the shower impacts the ground, and $r_0$ is the distance from the FD station to the shower core.
When the EAS geometry has been determined, the shower's longitudinal development is calculated by the inverse Monte Carlo method~\cite{bib:TA_IMC}.
This inverse Monte Carlo technique iteratively explores the longitudinal-development parameter space, searching for the optimum solution to reproduce the observed shower image.
After the detemination of longitudinal development, the calorimetric energy of the air shower, $E_{\rm{cal}}$, is evaluated by integrating the resulting Gaisser-Hillas function~\cite{bib:gh_function}.
Figure~\ref{fig:example} shows an example of a UHECR event reconstructed by the monocular analysis, comparing the signal observed at numerous points along the shower with the signal predicted by the best-fit profile reconstruction. The periodic structure of the profile is an artifact of the non-uniform sensitivity of the PMT photo-cathodes, an effect that is accurately simulated during reconstruction.

\subsection{Atmospheric model and fluorescence yield}
To infer the atmospheric energy deposit in the shower from the light measured by the detector, we must specify both a fluorescence yield and an atmospheric model, which respectively describe the production and attenuation of light prior to detection.
The atmospheric density profile is obtained from radiosonde pressure and temperature measurements in one-month averages at the city of Elko, Nevada, which is located $\sim$300~km to the northwest of the TA site.
The typical atmospheric transparency, determined by LIDAR operation~\cite{bib:lidar_tomida}, is characterized by a horizontal attenuation length of 29.4~km with a 1.0-km scale height, corresponding to a vertical aerosol optical depth (VAOD) of 0.034.
The absolute fluorescence yield is obtained from a measurement by Kakimoto~et~al.~\cite{bib:kakimoto_yield}, with a wavelength spectrum adopted from the result of the FLASH collaboration~\cite{bib:flash}.
The emission of fluorescence photons is proportional to the longitudinal projection of atmospheric energy deposit, with a lateral distribution described by the NKG function~\cite{bib:nkg1,bib:nkg2}.

\subsection{Air shower model}
We use the CORSIKA software~\cite{bib:corsika} to simulate the development of UHECRs using proton and iron primary particles, each according to five types of hadronic-interaction model: QGSJet01C, QGSJetII-03, QGSJetII-04, Sibyll 2.1, and Epos-LHC. 
CORSIKA also allows us to estimate the fraction of a primary's energy that is not deposited by charged particles in the air and does not contribute to the calorimetric energy. This missing energy for proton or iron primaries is parameterized by
\begin{equation}
\frac{E}{E_{\rm{cal}}} = a_{1} + a_{2}\log_{10}\frac{E_{\rm{cal}}}{\rm{eV}}+a_{3}\left(\log_{10}\frac{E_{\rm{cal}}}{\rm{eV}}\right)^2
\label{eq:missing_energy}
\end{equation}
where $E$ is the primary energy and $E_{\rm{cal}}$ is the calorimetric energy. Using the QGSJetII-03 model, the values ($a_1, a_2, a_3$) obtained for the missing energy are (3.083 $\pm$ 0.040, $-0.1947$ $\pm$ 0.0026, 0.00470 $\pm$ 0.00011) for proton, and (4.051 $\pm$ 0.024, $-0.2757$ $\pm$ 0.0016, 0.00639 $\pm$ 0.00007) for iron.
The estimated missing energy is 8\% -- 13\% for proton depending on energy; the difference between proton and iron is 7\% at $10^{17.5}$~eV and 4\% at $10^{19}$ eV.
Our missing-energy correction combines the proton and iron results, assuming the energy-dependent proton fraction as described below.

\begin{figure}[h]
    \centering
    \includegraphics[width=1\linewidth]{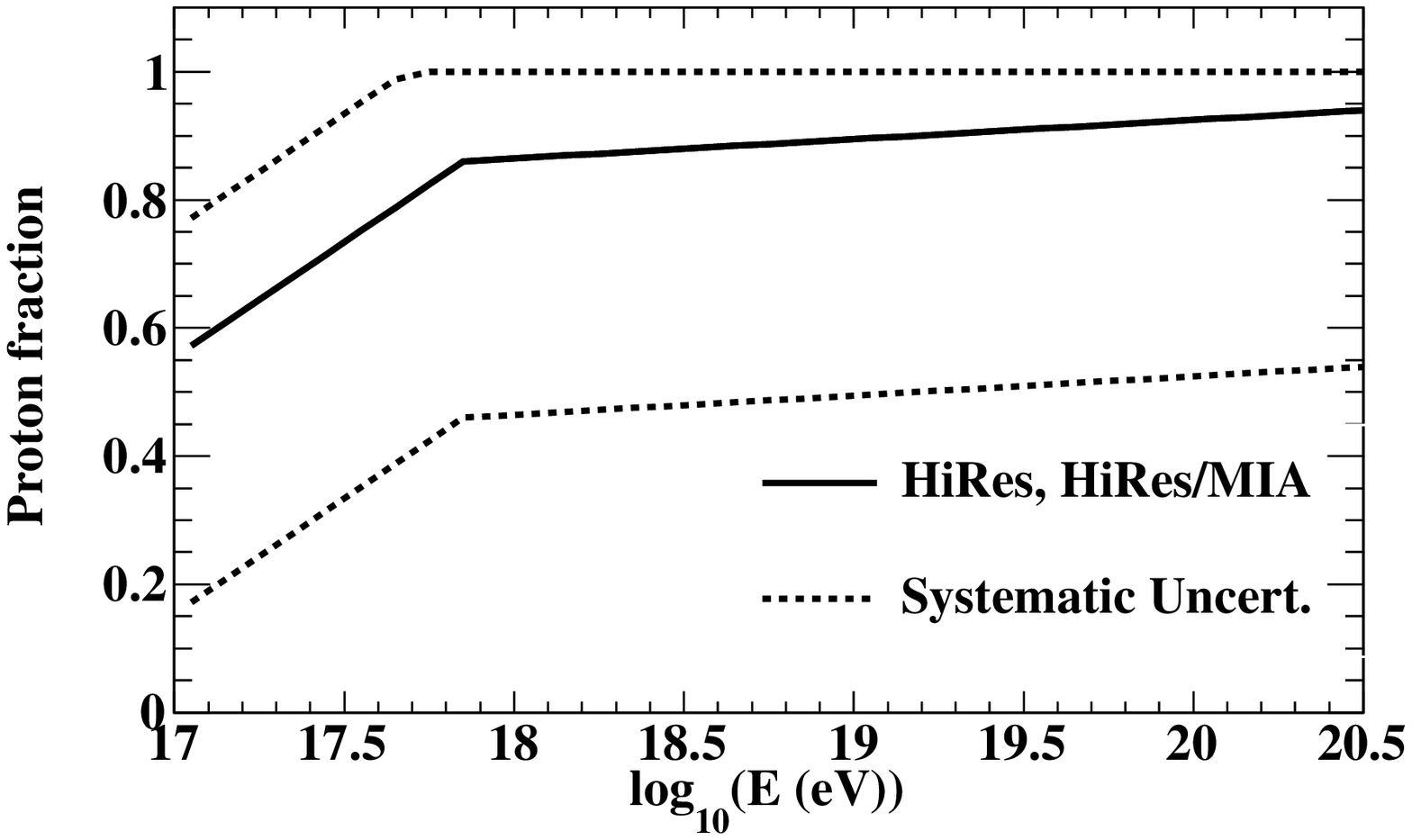}
    \caption{Proton fraction based on the result reported by the HiRes and HiRes/MIA experiment and its uncertainty~\cite{bib:hires_mia2005, bib:hires_mia}. The uncertainty of the proton fraction is indicated by the dashed line. If the fraction is larger than 1, purely proton is assumed.}
    \label{fig:proton_fraction}
\end{figure}

Figure~\ref{fig:proton_fraction} shows the proton fraction assumed in the FD analysis.
The proton fraction is obtained by fitting the HiRes and HiRes/MIA results~\cite{bib:hires_mia2005, bib:hires_mia}.
The HiRes and HiRes/MIA results indicate $\sim$50\% proton primaries at 10$^{17}$~eV, increasing to $\sim$90\% for energies above 10$^{18}$~eV. 
This assumed proton fraction is the same as that used in the HiRes-II spectrum~\cite{bib:hires_gzk}. 

The mass composition reported by the Pierre Auger Observatory suggests a transition from light nuclei at around $10^{18.3}$ eV to heavier nuclei up to energies of $10^{19.6}$ eV~\cite{bib:mass_auger, bib:mass_implication_auger, bib:auger_mass_icrc2015}.
However, the mass composition reported by both TA and Auger are consistent within systematic uncertainties~\cite{bib:composition_wg, bib:composition_wg2014}.
The mass composition at lower end of energy range is also reported from the KASCADE-Grande experiment~\cite{bib:kascade_light, bib:kascade_icrc2015}.
Therefore, we conservatively choose an uncertainty of the proton fraction to be +20\% and -40\% to accommodate their published results.

\subsection{Quality cuts}
The geometries of some showers, e.g., those that are too short or faint, are difficult to reconstruct accurately. 
Thus, we apply quality cuts to select only well reconstructed events in our analysis: the number of hit PMTs is larger than 10; the length of detected shower track in the FD FOV subtends an angle greater than 10$^{\circ}$ on the sky; the time extent is larger than 2~$\mu$s; and the depth of EAS maximum, $X_{\rm max}$, is within the FOV of the station, falling between the first and the last visible depths ($X_{\rm start}$ and $X_{\rm end}$, respectively). To avoid Cherenkov-light contamination, we require an angle on the shower-detector plane less than 120$^{\circ}$, and a minimum viewing angle greater than 20$^{\circ}$. 
Since our Monte Carlo (MC) simulations are generated with zenith angles up to 65$^\circ$ and core locations in a circle of 35-km radius centered on the CLF, the last two cuts are required to avoid contamination of the data with events from beyond the generated range, due to the monocular geometry resolution.
A full set of the quality cuts are summarized in Table~\ref{tbl:selection}.

\begin{table}
  \centering
  \begin{tabular}{lrr} \hline
    Quality Cuts                                 & events  & $f$ (\%) \\ \hline
    Reconstructed events                         & 86234   & ---  \\
    Number of PMTs $>10$                         & 85891   & 99.6\\
    Track length $>10^\circ$                     & 85738   & 99.8 \\
    Time extent $>2$~$\mu$s                      & 85452   & 99.7 \\
    $R_{p}$ $>0.5$ km                            & 83771   & 98.0 \\
    Minimum viewing angle $>20^\circ$            & 68560   & 81.8 \\
    $\Psi$ angle $<120^\circ$                    & 66504   & 97.0 \\
    Geometrical $\chi^2$/ndf $<10$               & 56247   & 84.6 \\
    $X_{\rm max}$ inside FOV                     & 30179   & 53.7 \\
    Zenith angle $<55^\circ$                     & 28959   & 96.0 \\
    Core distance from CLF $<25$ km              & 28932   & 99.9 \\ 
    Energy $>10^{17.2}$ eV                       & 28269   & 97.7 \\ \hline
    Total retention fraction                     & ---     & 37.8 \\ \hline
 \end{tabular}
\caption{Summary of the quality cuts. The number of events passing each successive selection criterion is described, together with the corresponding retention fraction $f$ of hitherto-passing events.}
 \label{tbl:selection}
\end{table}

\subsection{Resolution}
\begin{figure}[h]
  \centering
  \includegraphics[width=1.0\linewidth]{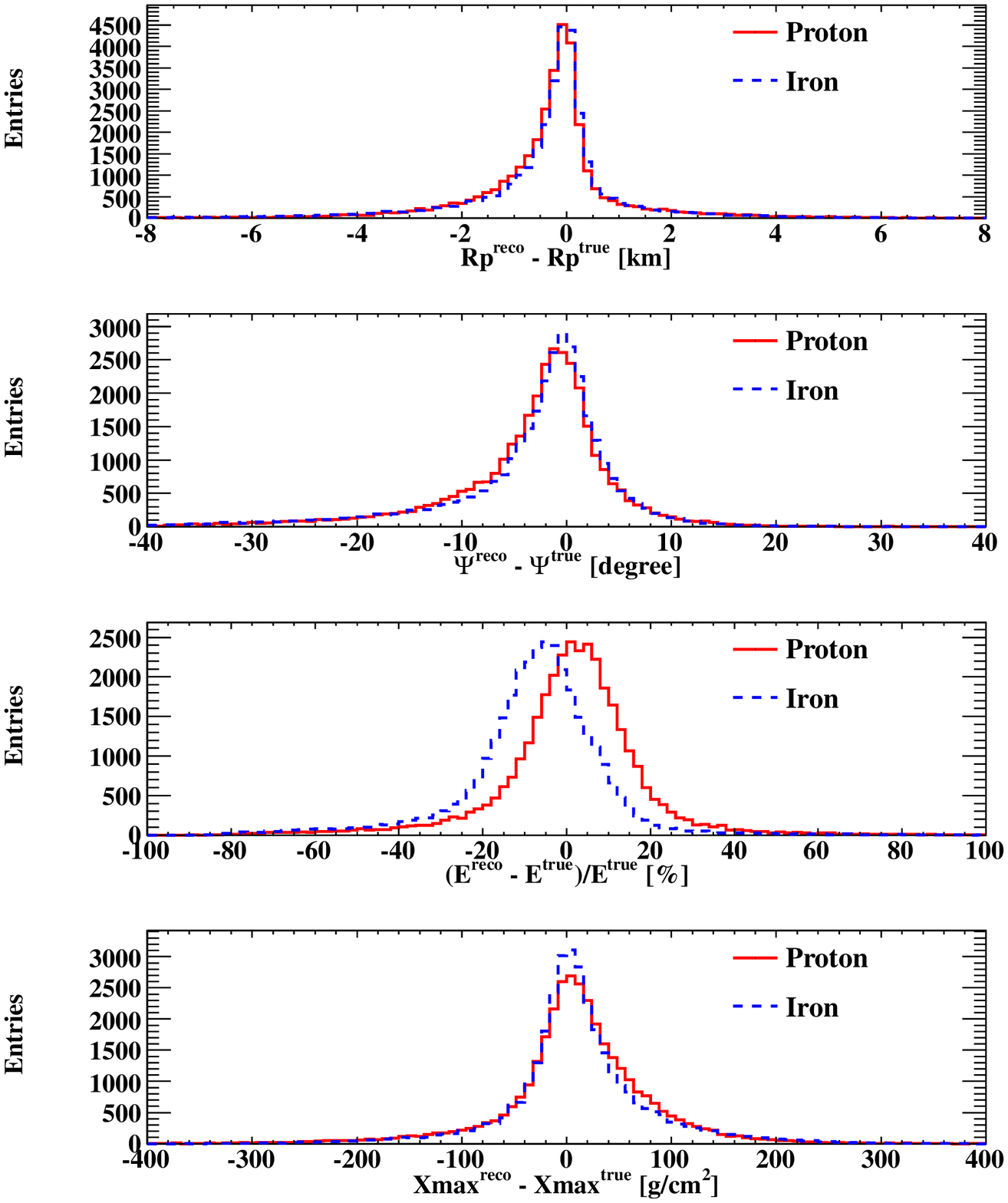}
  \caption{Resolutions of the monocular analysis estimated by the reconstruction of artificial data using proton (solid histogram) and iron (dashed histogram) with the QGSJetII-03 model.}
  \label{fig:resolution}
\end{figure}

The accuracy, or resolution, of the monocular analysis is estimated by reconstructing artificial data generated by Monte Carlo detector simulation, and comparing the reconstructed shower parameters to their true values as used by the simulation.
Figure~\ref{fig:resolution} shows the estimated resolution of the impact parameter $R_{p}$, angle on SDP $\Psi$, primary energy, and $X_{\rm max}$ for proton and iron primaries of the QGSJetII-03 interaction model.

As seen in Figure~\ref{fig:resolution}, the shapes of their histograms are the same between proton and iron primaries except for the reconstructed energy, because the missing energy is corrected assuming the proton fraction measured by HiRes and HiRes/MIA with the QGSJetII-03 model.
The systematic shift of +2\% on the reconstructed energy of proton-primary showers is derived from this missing energy correction.
The reconstructed energy of iron-primary showers is underestimated about 6\% because of the difference in the missing-energy fraction.
The asymmetric distributions in the $R_p$ and $\Psi$ resolution plots, seen as extended tails to the left of the peak, arise from the monocular PMT timing fit. The timing separation between consecutive PMTs is larger for ``receding'' showers ($\Psi<90^{\circ}$) than for incoming ones, leading to better resolution for the former than the latter. 
Typical resolutions of the monocular analysis are: 1.4~km in $R_{p}$, $7.7^\circ$ in $\Psi$ angle, 17\% in energy, and 72~g/cm$^{2}$ in $X_{\max}$.
The energy resolution is only weakly dependent on the primary energy, 16\% at 10$^{17.5}$~eV and 19\% at 10$^{19}$~eV.

\section{Aperture evaluation}
To evaluate the energy spectrum of UHECRs, it is essential to calculate the aperture and exposure of the FD stations.
The aperture can not be calculated as a simple geometrical factor because it depends on 
not only the energies, but also the primary species, as well as variability in FD 
performance, PMT sensitivity, and the atmosphere itself.
Thus, we estimate the FD aperture using MC simulations including these dependences.
The FD aperture, $A \Omega$, is calculated from the ratio between the number of 
reconstructed events passing the quality cuts and the number of thrown showers: $A \Omega (E) = A \Omega^{\rm G} \cdot N_{\rm reco}(E) / N_{\rm thrown}(E_{\rm thrown})$,
where $A \Omega^{\rm G}$ is the thrown-aperture region of MC simulation, $E_{\rm thrown}$ is the thrown energies, $N_{\rm reco}$ is the number of reconstructed events, and $N_{\rm thrown}$ is the number of thrown events.

\begin{figure}[h]
    \centering
    \includegraphics[width=1\linewidth]{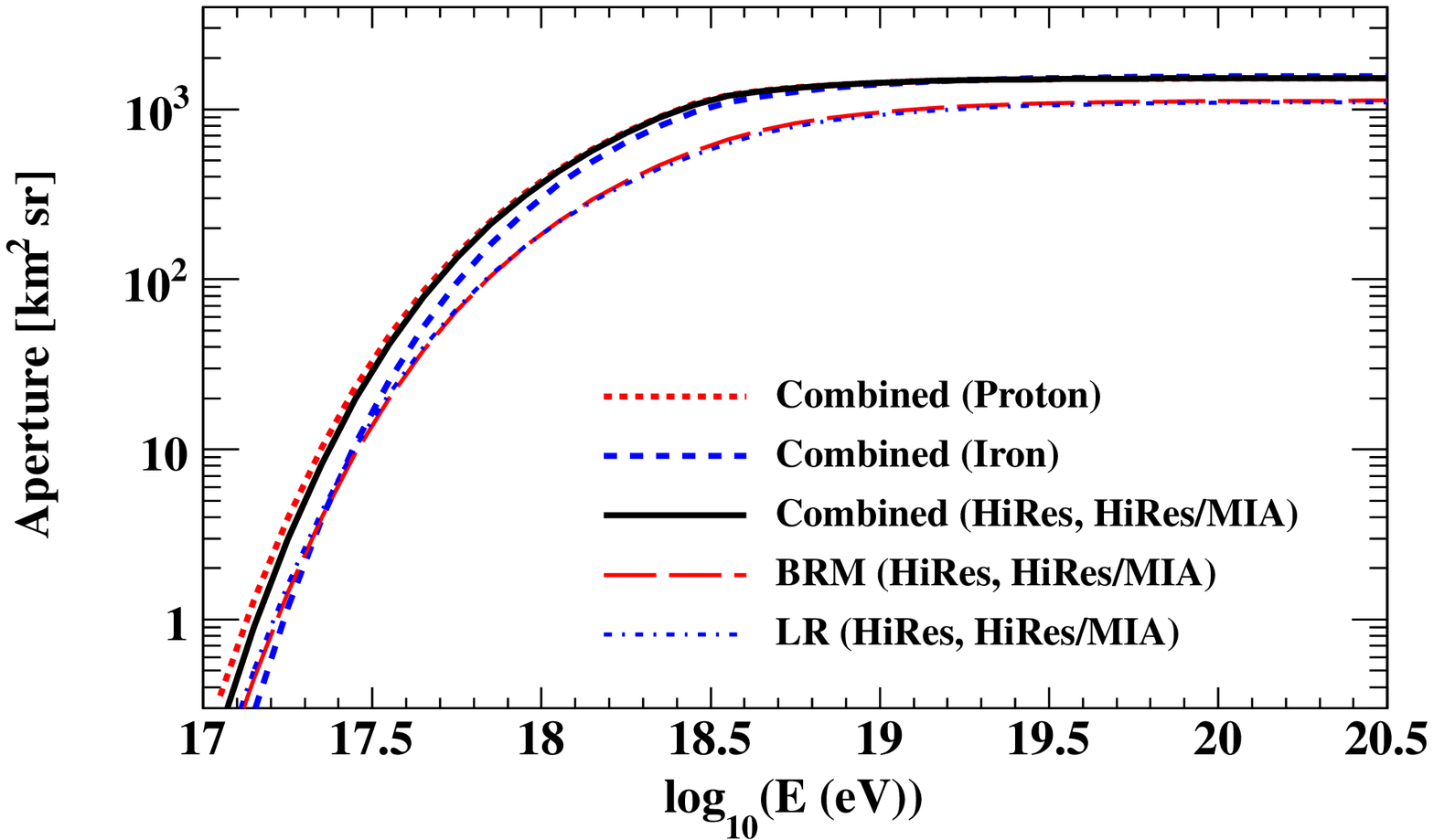}
    \caption{The combined aperture of the BRM and LR FD stations in monocular mode evaluated by MC simulations for primary compositions of the proton fraction measured by the HiRes and HiRes/MIA experiments (solid line), pure proton (dotted line) and pure iron (dashed line). The individual apertures at BRM and LR stations for the proton fraction by HiRes and HiRes/MIA experiments are indicated as long-dashed line and dash-dotted one.}
    \label{fig:aperture}
\end{figure}

Since TA was designed for stereoscopic observations of air showers at the higher energies, we define the combined aperture of the BRM and the LR stations in each monocular mode.
When an energetic shower is reconstructed by the both stations, we select whichever observation
has the larger number of detected photoelectrons to avoid any double-counting of high-energy showers.
Using the events reconstructed by both BRM and LR in their combined mode, we estimate the combined aperture of the BRM and LR stations with primary proton, iron nuclei, and the HiRes-and-HiRes/MIA reported composition as shown in Figure~\ref{fig:aperture}.
The obtained aperture (in units of km$^{2}$~sr), as calculated in each energy bin, is fit to a broken exponential function:
\begin{eqnarray}
\log_{10} A\Omega = p_1 \left( 1 - \exp \left( - \frac{\log_{10} E - p_2 }{ p_3 } \right) \right), E \le E_b \\
             = \gamma p_1 \left( 1 - \exp \left( - \frac{\log_{10} E - p_4 }{ p_5 } \right) \right), E \ge E_b 
\end{eqnarray}
where 
\begin{equation}
\gamma = \frac{ 1 - \exp \left( - \left(\log_{10} E_b - p_2 \right) / p_3  \right) }{ 1 - \exp \left( - \left(\log_{10} E_b - p_4 \right) / p_5  \right) }
\end{equation}
and $E_b$ is the energy (in eV) at the break. 
The fitting function is well reproduced on the aperture curve of FD and the best-fit values are described in Table~\ref{tbl:aperture_param}.
\begin{table}
  \centering
  \begin{tabular}{lrr} \hline
    Parameter     & Proton              & Iron \\ \hline
    $p_1$         &  3.55 $\pm$ 0.06    &  3.49 $\pm$  0.15 \\
    $p_2$         & 17.12 $\pm$ 0.01    & 17.22 $\pm$  0.01 \\
    $p_3$         &  0.68 $\pm$ 0.03    &  0.63 $\pm$  0.05 \\
    $p_4$         & 17.56 $\pm$ 0.07    & 17.40 $\pm$  0.13 \\
    $p_5$         &  0.29 $\pm$ 0.03    &  0.40 $\pm$  0.01 \\
    $\log_{10} E_b$    & 18.27 $\pm$ 0.09    & 17.87 $\pm$  0.03 \\  \hline
 \end{tabular}
\caption{The fit parameters for aperture assuming proton and iron primaries.}
\label{tbl:aperture_param}
\end{table}

The aperture assuming the proton fraction reported by HiRes and HiRes/MIA, $A\Omega^{f}$, was estimated by the following formula:
\begin{equation}
A\Omega^{f} = A\Omega^{\rm{P}} \left[ R + f \cdot \left(  1 - R  \right) \right],
\label{eq:fraction_aperture}
\end{equation}
where $f$ is the proton fraction and $R \equiv A\Omega^{\rm Fe}/A\Omega^{\rm P}$ is the ratio of the iron and proton best-fit apertures.
The dependence of the aperture on primary species is most evident in the low-energy
region, but becomes negligible at high energies.
These aperture calculations are based on QGSJetII-03, but the outcome does not strongly depend on the choice of hadronic interaction model: the model-to-model $X_{\rm max}$ variation is $\sim$25~g/cm$^2$, considerably smaller than the $\sim$100~g/cm$^2$ proton-iron separation shared by all models.

\section{Data analysis}
\begin{figure}[h]
   \includegraphics[width=1.0\linewidth]{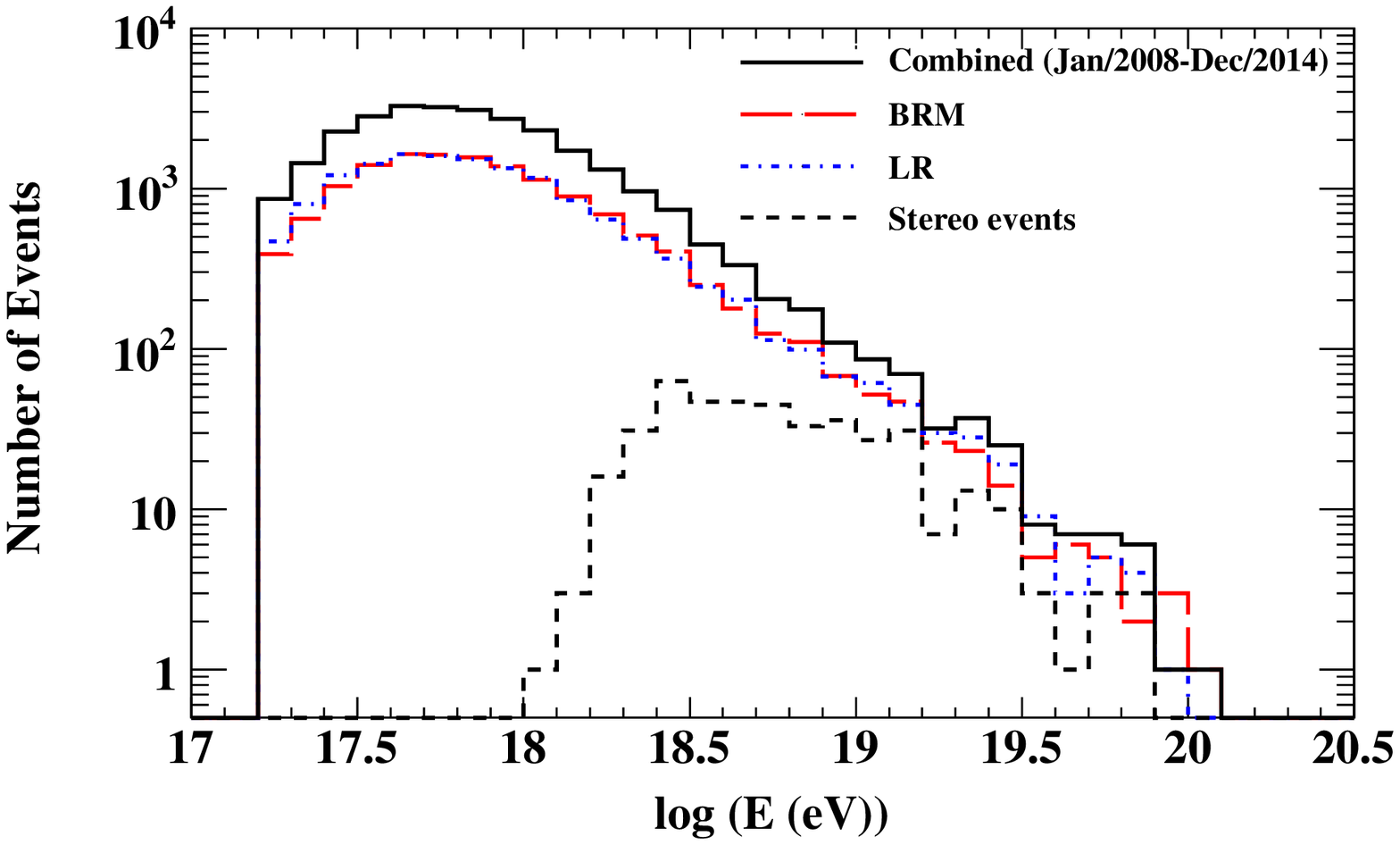}
\caption{Energy distribution of reconstructed showers from seven years of data. The dashed histogram shows events originally reconstructed by both stations, but with priority given to the result based on the larger number of detected photoelectrons to avoid duplication. The long-dashed and dash-dotted histograms show reconstructed showers at the BRM and LR stations.}
\label{fig:events}
\end{figure}

We analyze data collected by the BRM and LR FD stations from 1~January~2008 to 1~December~2014, corresponding to nearly seven years of observation.
The total live time (subtracting the dead time for data acquisition) is 6960 hours at BRM and 5850 hours at LR. 
A cloud cut ensures that we only analyze data collected under weather conditions 
that can be accurately modeled in our MC simulation. This cut is applied by interpreting the 
visually recorded code at the MD FD station because it has the most coverage in this period,
and we confirmed its consistency with the method described in Section~\ref{sec:ta}.
After the cloud cut, the live time is 4100 hours at BRM and 3470 hours at LR, 
so that 41\% of our data period was excluded by the cloud cut.
The live time of simultaneous BRM and LR observation is 2870 hours, with the remaining 1230 hours and 600 hours for the respective solo operation of BRM and LR.
Analyzing data using the monocular analysis under the same quality cuts, 28269 shower candidates above 10$^{17.2}$~eV are obtained as shown in Figure~\ref{fig:events}.
The number of events passing each selection criterion in sequence is summarized in Table~\ref{tbl:selection}. 
The energy distributions reconstructed at individual BRM and LR stations are also indicated in Figure~\ref{fig:events}; to avoid duplication when showers are reconstructed by both stations, we include only the result from whichever station detects more photoelectrons.

\begin{figure}[h]
   \includegraphics[width=1.0\linewidth]{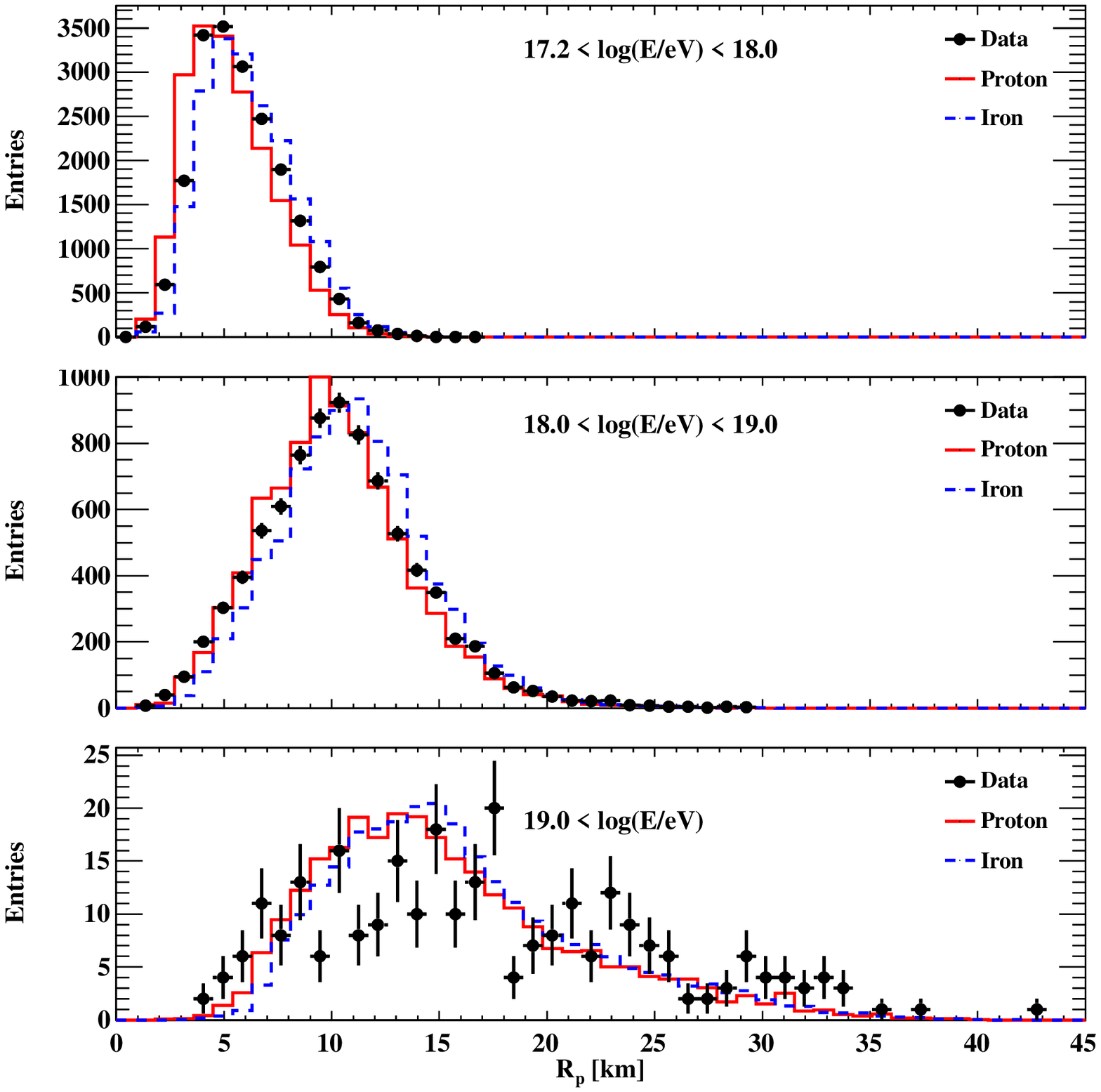}
   \caption{$R_p$ comparison in three energy ranges between observed data (points) and expectation estimated from the detector simulation using primary proton (solid histogram) and iron (dashed histogram) at the BRM and the LR stations. The histograms are weighted according to the energy spectrum measured by the surface detector array of the Telescope Array experiment \cite{bib:tasd_spectrum}. }
\label{fig:rp_datamc}
\end{figure}

\begin{figure}[h]
   \includegraphics[width=1.0\linewidth]{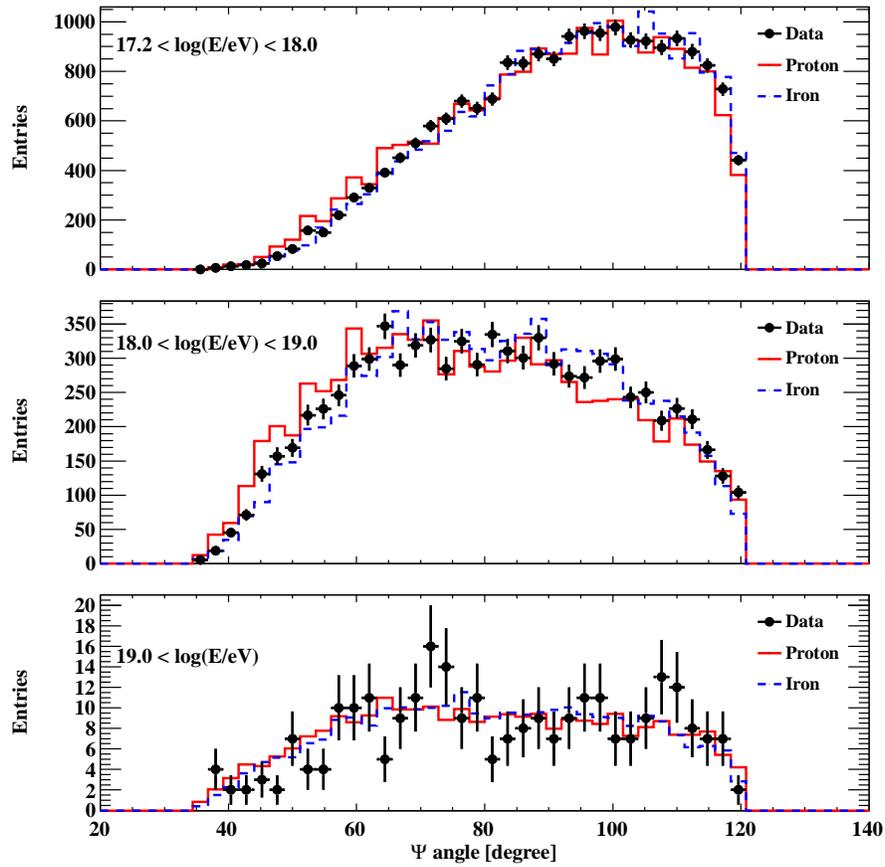}
  \caption{$\Psi$ comparison using the same method as in Figure~\ref{fig:rp_datamc}.}
\label{fig:psi_datamc}
\end{figure}

\begin{figure}[h]
  \includegraphics[width=1.0\linewidth]{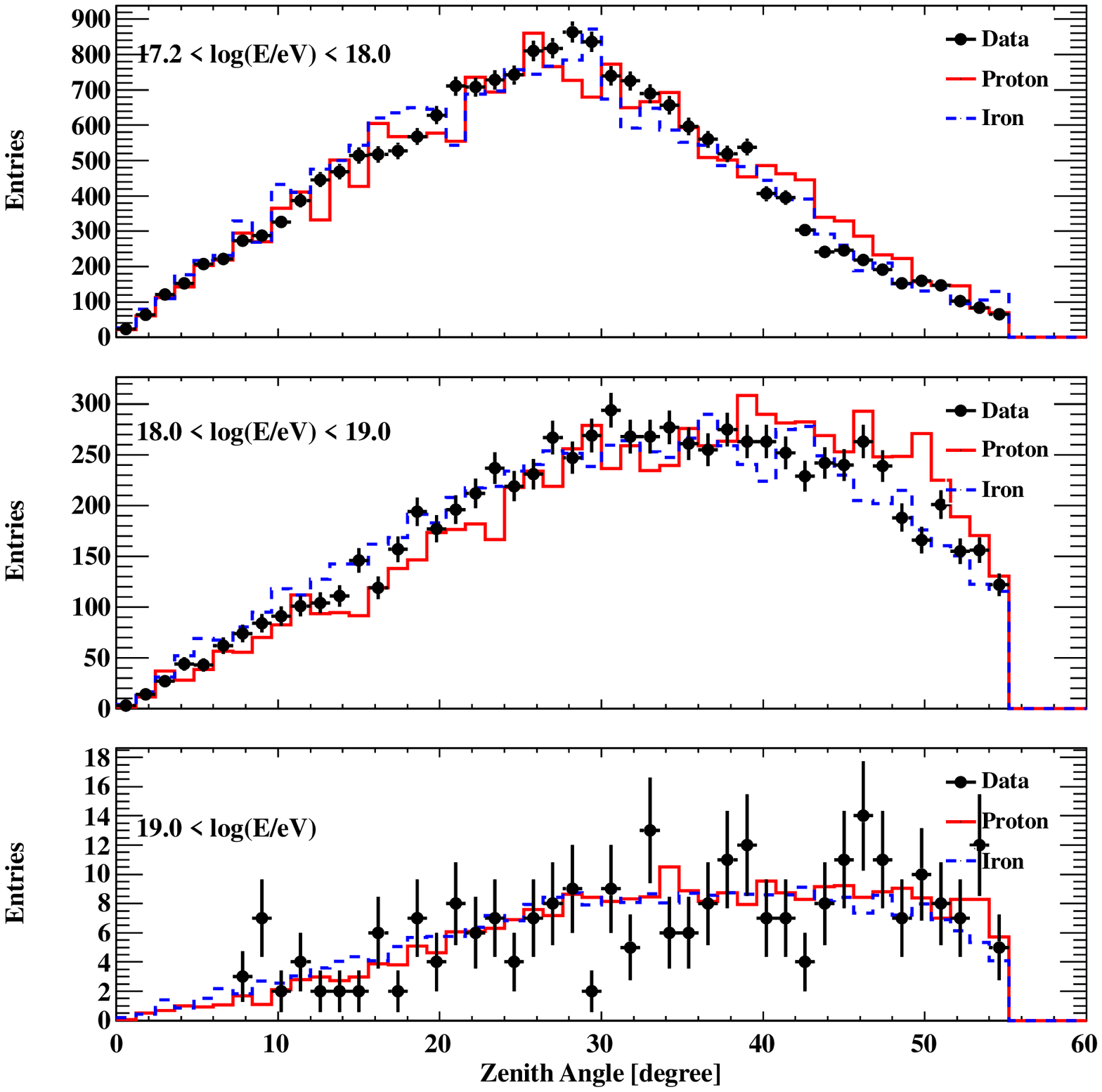}
  \caption{Zenith angle comparison using the same method as in Figure~\ref{fig:rp_datamc}.}
\label{fig:zen_datamc}
\end{figure}

\begin{figure}[h]
   \includegraphics[width=1.0\linewidth]{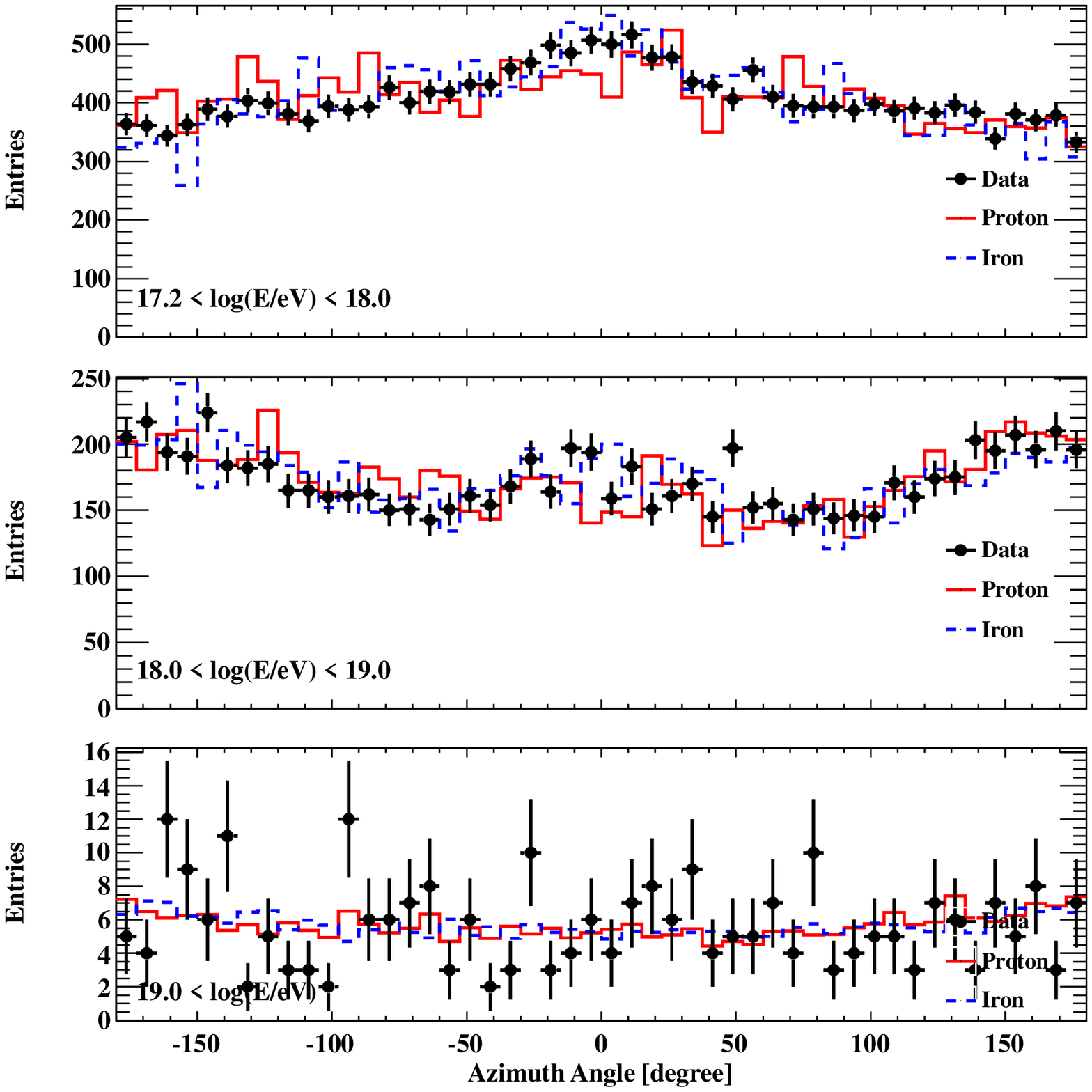}
  \caption{Azimuth angle comparison using the same method as in Figure~\ref{fig:rp_datamc}.}
\label{fig:azi_datamc}
\end{figure}

To further ensure the reliability of our analysis, the distributions of several parameters obtained from reconstruction of the observed data are compared with the predictions estimated from MC simulations using the QGSJetII-03 model.
The MC simulations are weighted according to the energy spectrum measured by the surface detectors~\cite{bib:tasd_spectrum}.
The comparisons of a variety of parameters are shown in Figures~\ref{fig:rp_datamc}--\ref{fig:azi_datamc} within three energy ranges: $10^{17.2-18.0}$~eV, $10^{18.0-19.0}$~eV, and above $10^{19.0}$~eV.
The distribution of each MC parameter is normalized to the number of data observations.

Above $10^{19}$ eV, the observed distribution is consistent with predictions estimated by either proton or iron, while at lower energies the observed distribution is bracketed by the predictions of proton and iron.
These distributions of observed data are what is expected if the mass composition in this energy range is a mixture of proton and iron as shown in Figure~\ref{fig:proton_fraction}, or intermediate nuclei.
The uncertainty on the mass composition has already been taken into account in the aperture calculation as shown in Figure~\ref{fig:aperture}. 
The apertures of proton and iron would provide bounds when estimating the energy spectrum.
The distribution could also be related with the dependence on hadronic interaction model, atmospheric model or fluorescence yield.
However, if we consider those uncetainties, those distributions shows a reasonable agreement between the observed data and the MC simulations.

\begin{figure}[h]
   \includegraphics[width=1.0\linewidth]{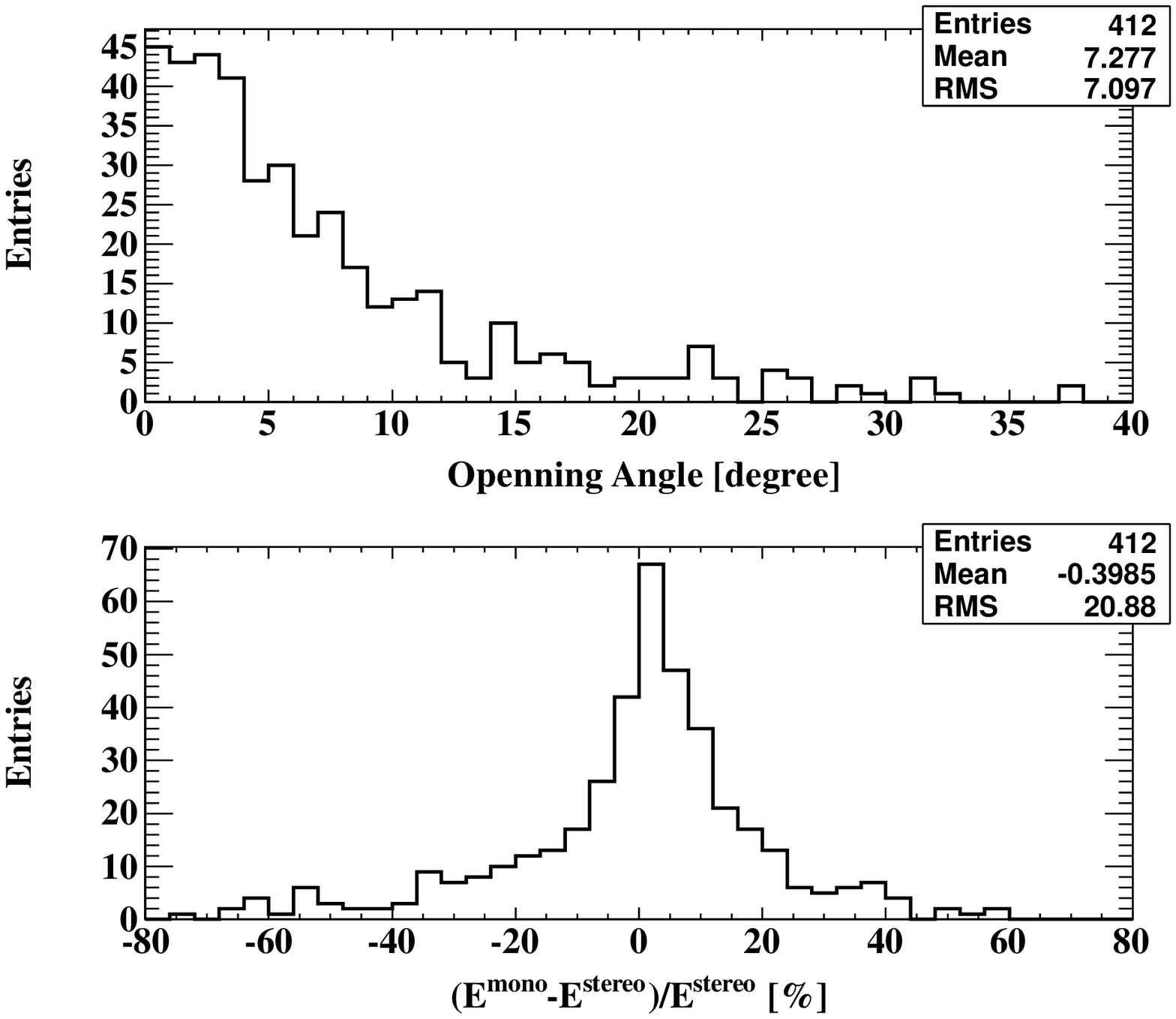}
  \caption{Comparison between monocular measurements and high-precision stereoscopic measurements of shower axis vector (top; opening angle) and reconstructed energy (fractional difference relative to calculation using stereoscopic geometry).}
\label{fig:mono_stereo}
\end{figure}

As another cross-check, we compared the results of monocular reconstruction with the high-precision measurements made possible by stereoscopic observation.
Figure~\ref{fig:mono_stereo} shows the distribution of differences between the stereoscopic and monocular analysis (for those events for which both measurements exist) in two representative quantities: the angle between reconstructed shower axis directions, and the fractional difference in reconstructed energies using the monocular and stereoscopic geometries.
These results are consistent with the expectation considering the resolution of the monocular analysis.

\subsection{Exposure calculation}
Since the BRM and LR stations with the common sensitive area are independently operated, some EASs with high energies (above $10^{18.5}$~eV) are reconstructed at both stations. 
Thus, we carefully select results of both reconstructed EASs to avoid any double-counting of EASs.
As a result, the reconstructed EASs include results reconstructed from three types of observation mode: the respective monocular modes of the BRM or LR station operating alone, and the simultaneous operation of both.
Therefore, considering the three types of the apertures and the live times, we estimate the total exposure, $\mathcal{E}$. 
The exposures of the BRM and the LR stations are simply given by
\begin{equation}
\mathcal{E}(E)= A\Omega(E)  \cdot t,
\end{equation}
where $t$ indicates the live time, and the aperture $A\Omega(E)$ is already estimated in Figure~\ref{fig:aperture}.
We consider the three types of the apertures and the three types of the live times: BRM only, LR only, and the simultaneous operation of both.
Thus, the total exposure is modified as
\begin{equation}
\mathcal{E}^{\rm total}(E) = \sum_{i={\rm brm,lr,both}} A\Omega^{i}(E) \cdot t^{i}.
\label{eq:total_exposure}
\end{equation}

\subsection{Energy spectrum}
The flux measured at the BRM and the LR stations is straightforward to evaluate with the following formula:
\begin{equation}
J(E) = \frac{N(E)}{\mathcal{E}^{\rm total}(E) \cdot \Delta E },
\label{eq:energy_spectrum}
\end{equation}
where $N(E)$ is the number of events in the each bin and $\Delta E$ is the width of each bin.

\begin{figure}[h]
    \centering
    \includegraphics[width=1\linewidth]{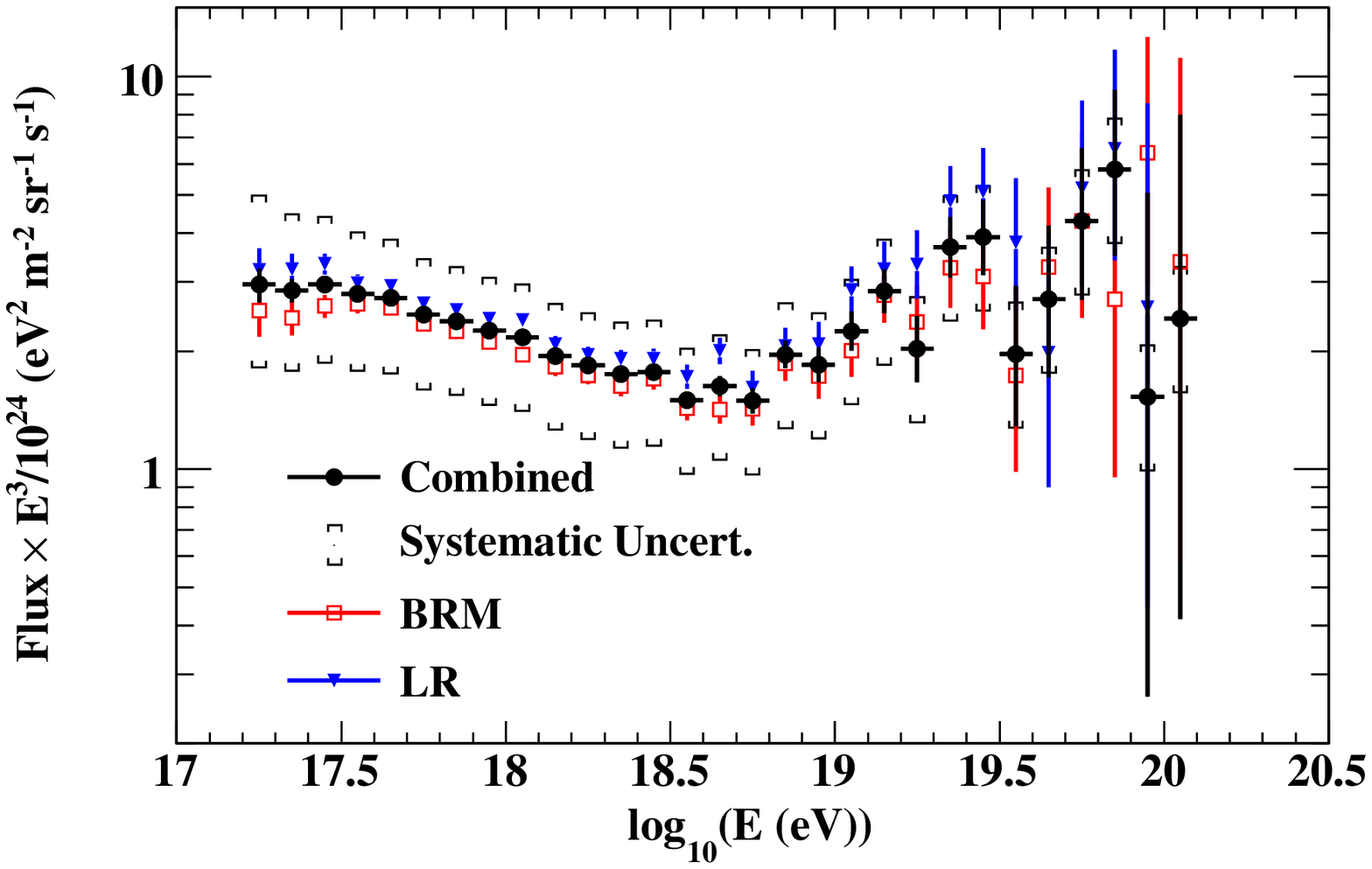}
    \caption{Energy spectra observed by BRM and LR separately, and combined. The total systematic uncertainty on flux to be discussed in Section~\ref{sec:uncertainty} is also indicated.}
    \label{fig:spectrumBRMLR}
\end{figure}

The energy spectra obtained separately by BRM and LR, and by both in combination, are shown in Figure~\ref{fig:spectrumBRMLR}.
The BRM and LR individual spectra suggest a different on energy scale by 5\%. 
The difference is explained by the systematic uncertainty on calibration and atmospheric model to be discussed in Section~\ref{sec:uncertainty}.
Since an arrangement of BRM and LR stations has a common sensitive area, some showers at high energies are reconstructed by both stations.
In the combined spectrum, the reconstructed shower is selected by choosing the one with the larger number of detected photoelectrons. This avoids duplication in counter showers. This also implies that the combined spectrum is not located between the two individual spectra at high energies.

\begin{figure}[h]
    \centering
    \includegraphics[width=1\linewidth]{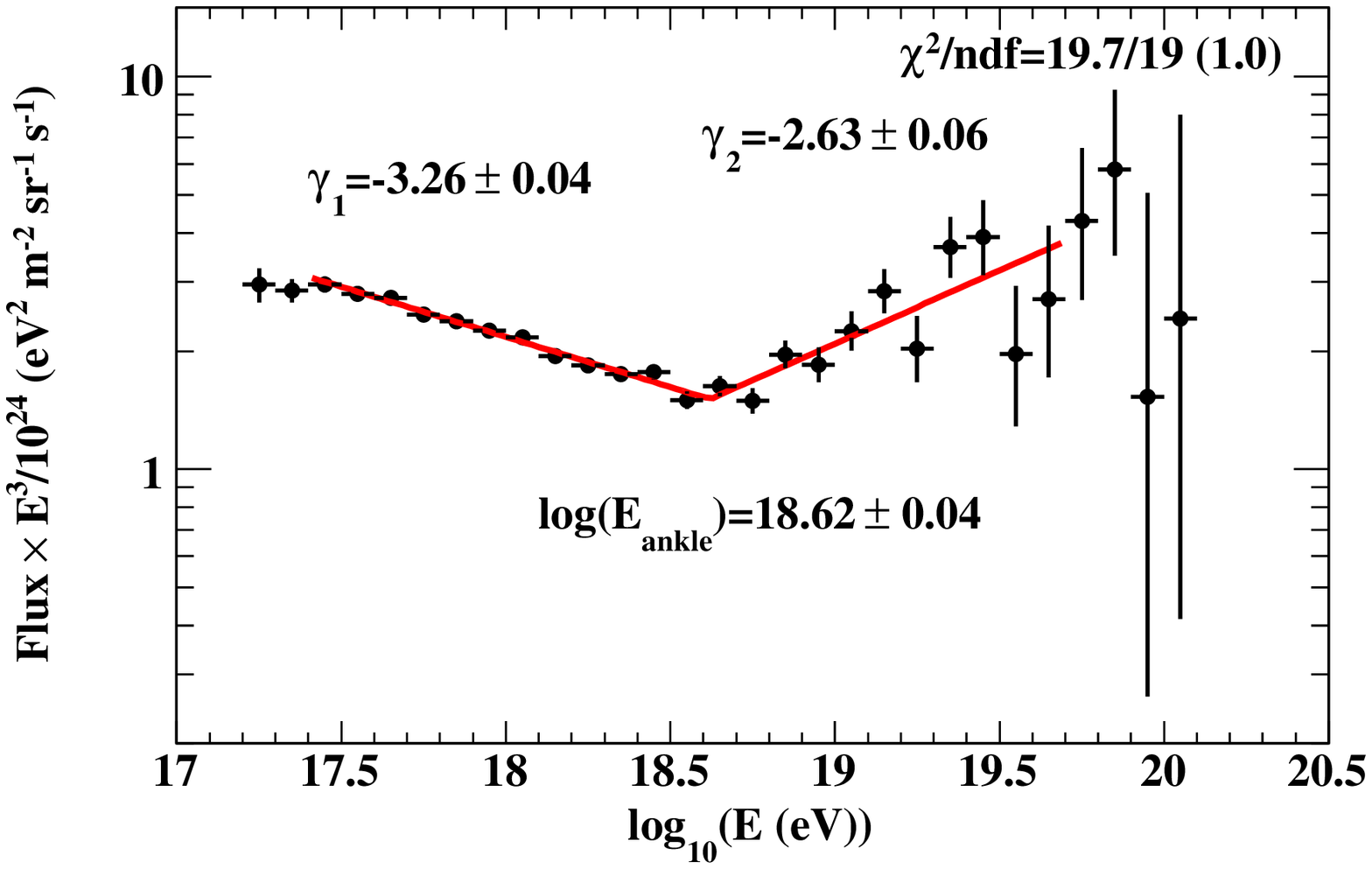}
    \caption{Fitted result on the combined energy spectrum observed by the BRM and LR fluorescence detector stations.}
    \label{fig:spectrum}
\end{figure}

The spectrum with a broken-power-law fit is shown in Figure~\ref{fig:spectrum}.
The fitted region was specified below the suppression energy from the TA SD report~\cite{bib:tasd_spectrum}. 
There is an obvious break at $\log_{10}(E_{\rm{ankle}}) = 18.62 \pm 0.04$, corresponding to the ankle. 
The spectral index is $3.26 \pm 0.04$ below the break and $2.63 \pm 0.06$ above it.
The suppression observed by the TA SD array above 10$^{19.7}$~eV is less significant here because of statistical limitations.

\begin{figure}[h]
    \centering
    \includegraphics[width=1\linewidth]{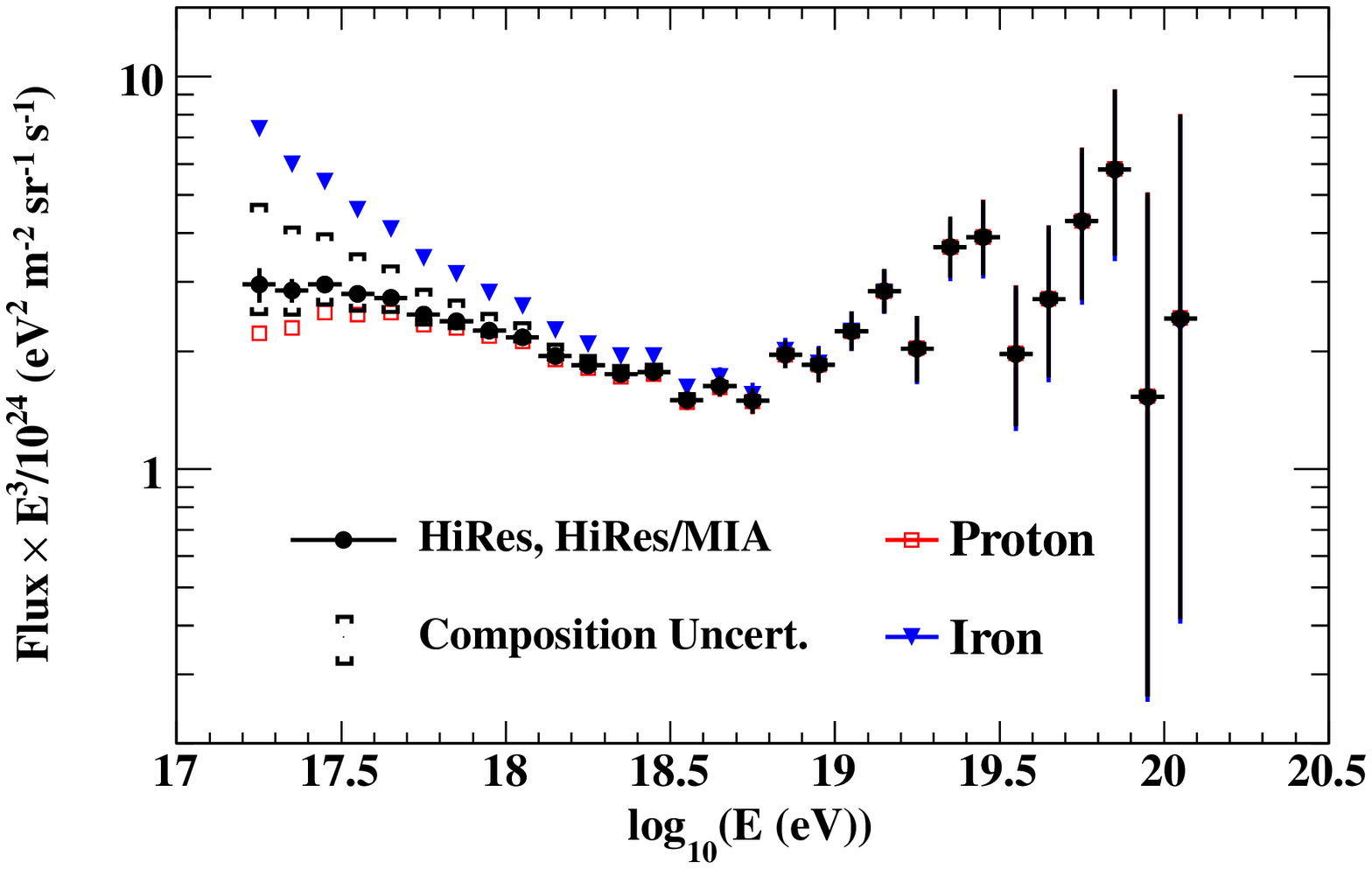}
    \caption{Systematic uncertainty attributed to the assumed proton fraction. The open-squares are the flux estimated by an aperture assuming primary proton, and the triangles assume primary iron. The square bracket corresponds to the uncertainty caused by +20\% and -40\% uncertainty on the proton fraction. }
    \label{fig:spectrum_uncert}
\end{figure}

To evaluate the energy spectrum, we use the aperture calculated assuming the proton 
fraction reported by the HiRes and HiRes/MIA experiments.
We choose +20\% and -40\% uncertainty of the proton fraction, and then estimate the aperture by Equation~\ref{eq:fraction_aperture}. 
Since the difference between hadron interaction models is negligible for the aperture calculation compared with the effect of the proton fraction,
we choose the QGSJetII-03 model, which shows a good agreement with other TA analyses.
Figure~\ref{fig:spectrum_uncert} shows the uncertainty of the energy spectrum attributable to aperture evaluation assuming this proton fraction.
For reference of the bounding estimation, the energy spectra assuming purely proton and iron compositions are also indicated.
In the lower-energy region, the energy spectrum is dependent on the assumed proton fraction, while this dependence is negligible in the higher-energy region. 

\section{Systematic uncertainty on the energy scale}
\label{sec:uncertainty}
In TA FD analysis, we adopt the absolute fluorescence-photon yield measured by Kakimoto et al.~\cite{bib:kakimoto_yield}, and we also adopt the relative intensity spectrum from FLASH measurements~\cite{bib:flash}. Combining the respective uncertainties on these measurements produces an 11\% contribution to our own systematic uncertainty in measured energy.

In the atmospheric model, we use a median value of atmospheric transparency (VAOD = 0.034) for all seasons, and for the atmospheric pressure and temperature profiles we adopt the radiosonde measured at the Elko, Nevada airport in one-month averages.
The uncertainty of aerosol transparency is estimated from the standard deviation of VAOD LIDAR measurements, approximately 0.015, so that the uncertainty has an energy dependence: 3\% at $10^{17.2}$~eV and 10\% above $10^{19.5}$~eV. 
The uncertainty from the monthly radiosonde is a contribution of 5\%. 
Therefore the total systematic uncertainty contribution of the atmospheric model is conservatively chosen as 11\%.

The uncertainty caused by detector calibration is largly dependent on the absolute calibration using CRAYS, which is estimated as 8\% on energy scale~\cite{bib:CRAYS}.
The uncertainty of long-term variation in PMT gains is determined to be less than 3\% by analysis of PMTs equipped with YAP pulsers~\cite{bib:yap}.
Two additional constituents of systematic uncertainty are the uncertainties in mirror reflectance and the filter transmittance, and also their time variations; their respective contributions are 3\% and 1\%~\cite{bib:htokuno_ta_calib}.
The systematic uncertainty attributed to the uncertainties on telescope and PMT pointing directions are estimated as 4\%.
By adding these detector-calibration uncertainties in quadrature,
the total uncertainty attributed to the uncertainties on the detector calibrations is estimated to be 10\%.

Since the missing energy is corrected assuming the proton fraction measured by the HiRes and HiRes/MIA experiments in our reconstruction, this systematic uncertainty is evaluated as 4\%. 
Compared with results by an independently developed analysis, we confirmed the effect on the energy scale is less than 8\% in the relevant energy range~\cite{bib:sean_mc}.
The total uncertainty on reconstruction is estimated as 9\% by quadratic sum of those two components.

Adding all of the aforementioned uncertainties in quadrature, we conclude that the total systematic uncertainty on the energy scale is 21\%. 
When considering the power-law energy dependence of the spectrum, a 21\% uncertainty on energy scale turns into a 35\% uncertainty on the measurement of UHECR flux. 

\begin{figure*}[h]
    \centering
    \includegraphics[width=1\linewidth]{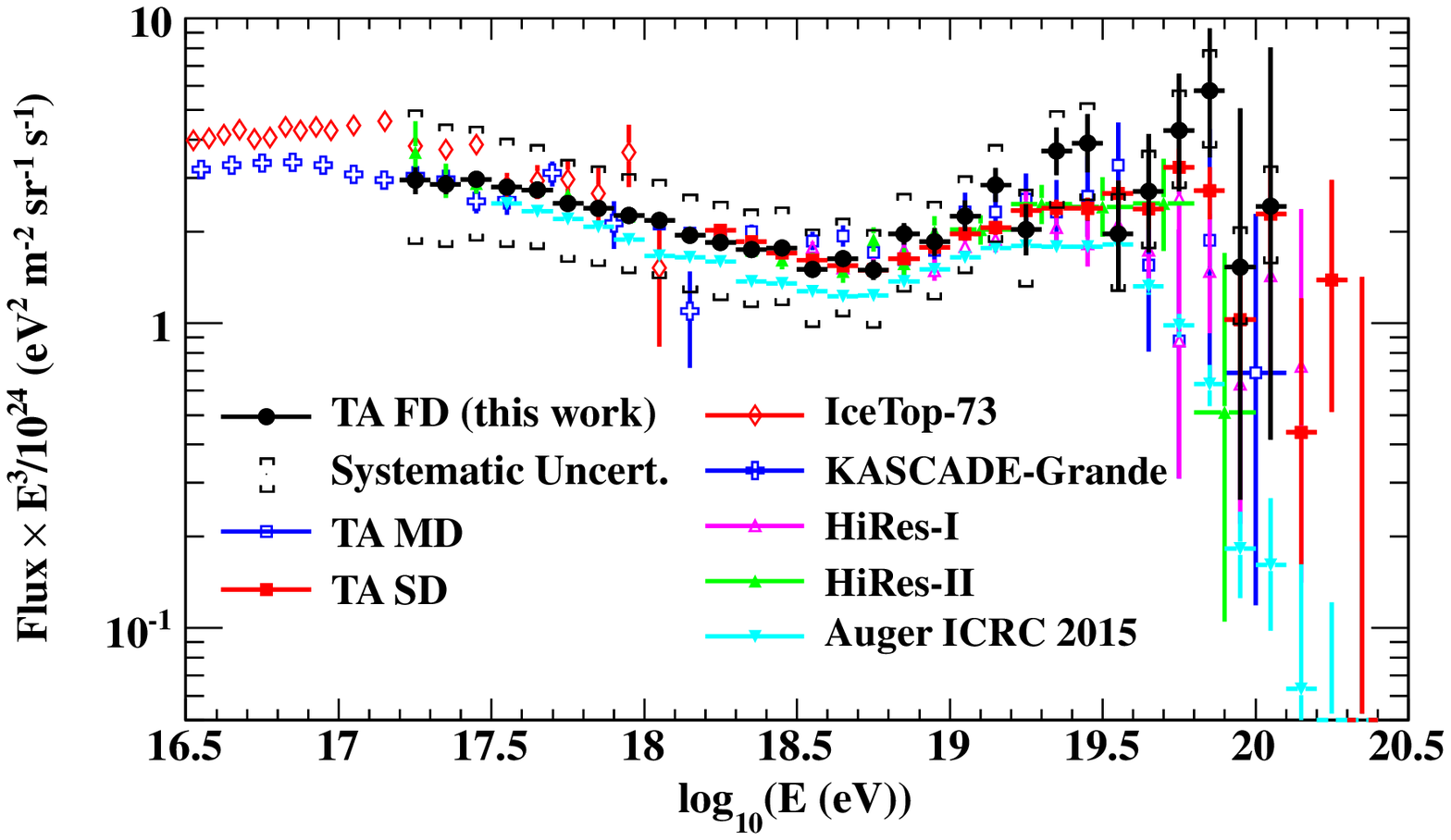}
    \caption{Energy spectrum compared with results reported by IceTop-73~\cite{bib:icetop-73}, KASCADE-Grande~\cite{bib:kascade_spectrum}, HiRes~\cite{bib:hires_gzk}, Auger~\cite{bib:AugerSDSpectrum} and other detectors within TA~\cite{bib:tasd_spectrum,bib:tamd_spectrum}. }
    \label{fig:energy_spectrum}
\end{figure*}

We can compare the obtained energy spectrum with other spectrum measurements reported by IceTop-73~\cite{bib:icetop-73}, KASCADE-Grande~\cite{bib:kascade_spectrum}, HiRes~\cite{bib:hires_gzk}, the Pierre Auger Observatory~\cite{bib:AugerSDSpectrum} and other detectors within TA~\cite{bib:tasd_spectrum,bib:tamd_spectrum}. 
As seen in Figure~\ref{fig:energy_spectrum}, our energy spectrum is in agreement with results reported from IceTop-73 and KASCADE-Grande within the systematic uncertainty.
As shown in the high energy range, the structure of the spectrum is in good agreement with the spectra reported using the TA surface detector and by HiRes-II.
Although the Auger spectrum is shifted 9\% lower in energy scale than our spectrum, it is also consistent within the systematic uncertainty on the energy scale.

In the case where we adopt the fluorescence yield reported by the AirFly experiment~\cite{bib:airfly1, bib:airfly2} which is used by the Auger experiment, the TA energy scale goes down by 14\%.
Therefore, the TA energy scale would be change to be 5\% lower than the Auger if we use the same fluroescence yield. 
This is within the systematic uncertainty.

\section{Conclusions}
\label{sec:conclusions}
We have measured the cosmic-ray energy spectrum covering three orders of magnitude at energies above $10^{17.2}$~eV using the monocular analysis of data taken during the first seven years of operation by the fluorescence detectors of the Telescope Array experiment.
The obtained spectrum has an overall broken-power-law structure, with an obvious spectral-index break at an energy of $\log_{10}(E_{\rm ankle}/{\rm eV}) = 18.62 \pm 0.04$, corresponding to the ankle. 
The structure is in good agreement with the spectra reported using the TA surface detectors and by HiRes-II.

\section*{Acknowledgments} 
The Telescope Array experiment is supported by the Japan
Society for the Promotion of Science through Grants-in-Aid for Scientific Research on Specially 
Promoted Research (21000002) ``Extreme Phenomena in the Universe Explored
by Highest Energy Cosmic Rays'' and for Scientific Research
 (19104006), and the Inter-University Research Program of
the Institute for Cosmic Ray Research; by the U.S. National
Science Foundation awards PHY-0307098, PHY-0601915, PHY-0649681,
PHY-0703893, PHY-0758342, PHY-0848320, PHY-1069280, PHY-1069286, PHY-1404495 and PHY-1404502; by the National Research Foundation
of Korea (2007-0093860, 2012R1A1A2008381, 2013004883); by
the Russian Academy of Sciences, RFBR grants 11-02-01528a and 13-02-01311a (INR), IISN project No. 4.4502.13; and Belgian
Science Policy under IUAP VII/37 (ULB). The foundations
of Dr. Ezekiel R. and Edna Wattis Dumke, Willard
L. Eccles, and George S. and Dolores Dor\'e Eccles all
helped with generous donations. The State of Utah supported
the project through its Economic Development Board, and the
University of Utah through the Office of the Vice President
for Research. The experimental site became available through
the cooperation of the Utah School and Institutional Trust
Lands Administration (SITLA), U.S. Bureau of Land Management,
and the U.S. Air Force. We also wish to thank the people
and the officials of Millard County, Utah for their steadfast
and warm support. We gratefully acknowledge the contributions
from the technical staffs of our home institutions. An
allocation of computer time from the Center for High Performance
Computing at the University of Utah is gratefully acknowledged.

\bibliography{SpectrumFDMono15}

\begin{thebibliography}{10}
\expandafter\ifx\csname url\endcsname\relax
  \def\url#1{\texttt{#1}}\fi
\expandafter\ifx\csname urlprefix\endcsname\relax\def\urlprefix{URL }\fi
\expandafter\ifx\csname href\endcsname\relax
  \def\href#1#2{#2} \def\path#1{#1}\fi

\bibitem{bib:history_cosmicray}
K.-H. Kampert, A.~A. Watson, {Extensive Air Showers and Ultra High-Energy
  Cosmic Rays: A Historical Review}, Eur.Phys.J. H37 (2012) 359--412.
\newblock \href {http://arxiv.org/abs/1207.4827} {\path{arXiv:1207.4827}},
  \href {http://dx.doi.org/10.1140/epjh/e2012-30013-x}
  {\path{doi:10.1140/epjh/e2012-30013-x}}.

\bibitem{bib:gzk1}
K.~Greisen, {End to the cosmic ray spectrum?}, Phys.Rev.Lett. 16 (1966)
  748--750.
\newblock \href {http://dx.doi.org/10.1103/PhysRevLett.16.748}
  {\path{doi:10.1103/PhysRevLett.16.748}}.

\bibitem{bib:gzk2}
G.~Zatsepin, V.~Kuzmin, {Upper limit of the spectrum of cosmic rays}, JETP
  Lett. 4 (1966) 78--80.

\bibitem{bib:TA}
M.~Fukushima, et~al., {Telescope Array project for extremely high energy cosmic
  rays}, Prog.Theor.Phys.Suppl. 151 (2003) 206--210.
\newblock \href {http://dx.doi.org/10.1143/PTPS.151.206}
  {\path{doi:10.1143/PTPS.151.206}}.

\bibitem{bib:TA_SD}
T.~Abu-Zayyad, et~al., {The surface detector array of the Telescope Array
  experiment}, Nucl.Instrum.Meth. A689 (2012) 87--97.
\newblock \href {http://arxiv.org/abs/1201.4964} {\path{arXiv:1201.4964}},
  \href {http://dx.doi.org/10.1016/j.nima.2012.05.079}
  {\path{doi:10.1016/j.nima.2012.05.079}}.

\bibitem{bib:htokuno_telescope}
H.~Tokuno, Y.~Tameda, M.~Takeda, K.~Kadota, D.~Ikeda, et~al., {New air
  fluorescence detectors employed in the Telescope Array experiment},
  Nucl.Instrum.Meth. A676 (2012) 54--65.
\newblock \href {http://arxiv.org/abs/1201.0002} {\path{arXiv:1201.0002}},
  \href {http://dx.doi.org/10.1016/j.nima.2012.02.044}
  {\path{doi:10.1016/j.nima.2012.02.044}}.

\bibitem{bib:hires}
J.~Boyer, B.~Knapp, E.~Mannel, M.~Seman, {FADC-based DAQ for HiRes Fly's Eye},
  Nucl.Instrum.Meth. A482 (2002) 457--474.
\newblock \href {http://dx.doi.org/10.1016/S0168-9002(01)01517-0}
  {\path{doi:10.1016/S0168-9002(01)01517-0}}.

\bibitem{bib:tasd_spectrum}
T.~Abu-Zayyad, et~al., {The Cosmic Ray Energy Spectrum Observed with the
  Surface Detector of the Telescope Array Experiment}, Astrophys.J. 768 (2013)
  L1.
\newblock \href {http://arxiv.org/abs/1205.5067} {\path{arXiv:1205.5067}},
  \href {http://dx.doi.org/10.1088/2041-8205/768/1/L1}
  {\path{doi:10.1088/2041-8205/768/1/L1}}.

\bibitem{bib:md_spectrum}
T.~Abu-Zayyad, et~al., {The Energy Spectrum of Telescope Array's Middle Drum
  Detector and the Direct Comparison to the High Resolution Fly's Eye
  Experiment}, Astropart.Phys. 39-40 (2012) 109--119.
\newblock \href {http://arxiv.org/abs/1202.5141} {\path{arXiv:1202.5141}},
  \href {http://dx.doi.org/10.1016/j.astropartphys.2012.05.012}
  {\path{doi:10.1016/j.astropartphys.2012.05.012}}.

\bibitem{bib:tafd_spectrum}
T.~Abu-Zayyad, et~al., {The Energy Spectrum of Ultra-High-Energy Cosmic Rays
  Measured by the Telescope Array FADC Fluorescence Detectors in Monocular
  Mode}, Astropart.Phys. 48 (2013) 16--24.
\newblock \href {http://arxiv.org/abs/1305.6079} {\path{arXiv:1305.6079}},
  \href {http://dx.doi.org/10.1016/j.astropartphys.2013.06.007}
  {\path{doi:10.1016/j.astropartphys.2013.06.007}}.

\bibitem{bib:ytameda_trigger}
Y.~Tameda, A.~Taketa, J.~D. Smith, M.~Tanaka, M.~Fukushima, et~al., {Trigger
  electronics of the new fluorescence detectors of the Telescope Array
  experiment}, Nucl.Instrum.Meth. A609 (2009) 227--234.
\newblock \href {http://dx.doi.org/10.1016/j.nima.2009.07.093}
  {\path{doi:10.1016/j.nima.2009.07.093}}.

\bibitem{bib:CRAYS}
S.~Kawana, N.~Sakurai, T.~Fujii, M.~Fukushima, N.~Inoue, et~al., {Calibration
  of Photomultiplier Tubes for the Fluorescence Detector of Telescope Array
  Experiment using a Rayleigh Scattered Laser Beam}, Nucl.Instrum.Meth. A681
  (2012) 68--77.
\newblock \href {http://arxiv.org/abs/1202.1934} {\path{arXiv:1202.1934}},
  \href {http://dx.doi.org/10.1016/j.nima.2012.03.011}
  {\path{doi:10.1016/j.nima.2012.03.011}}.

\bibitem{bib:htokuno_ta_calib}
H.~Tokuno, Y.~Murano, S.~Kawana, Y.~Tameda, A.~Taketa, et~al., {On site
  calibration for new fluorescence detectors of the Telescope Array
  experiment}, Nucl.Instrum.Meth. A601 (2009) 364--371.
\newblock \href {http://dx.doi.org/10.1016/j.nima.2008.12.210}
  {\path{doi:10.1016/j.nima.2008.12.210}}.

\bibitem{bib:yap}
B.~K.~Shin, H.~Tokuno, Y.~Tsunesada, T.~Abu-Zayyad, R.~Aida, et~al., {Gain
  monitoring of Telescope Array photomultiplier cameras for the first 4 years
  of operation}, Nucl.Instrum.Meth. A768 (2014) 96--103.
\newblock \href {http://dx.doi.org/10.1016/j.nima.2014.09.059}
  {\path{doi:10.1016/j.nima.2014.09.059}}.

\bibitem{bib:ta_clf}
S.~Udo, M.~Allen, R.~Cady, M.~Fukushima, Y.~Iida, J.~N.~Matthews, J.~Thomas,
  S.~B.~Thomas, L.~R.~Wiencke, {The Central Laser Facility at the Telescope
  Array}, Proc. of the 30th International Cosmic Ray Conference, Merida, Mexico
  5 (2007) 1021--1024.

\bibitem{bib:lidar_tomida}
T.~Tomida, Y.~Tsuyuguchi, T.~Arai, T.~Benno, M.~Chikawa, et~al., {The
  atmospheric transparency measured with a LIDAR system at the Telescope Array
  experiment}, Nucl.Instrum.Meth. A654 (2011) 653--660.
\newblock \href {http://arxiv.org/abs/1109.1196} {\path{arXiv:1109.1196}},
  \href {http://dx.doi.org/10.1016/j.nima.2011.07.012}
  {\path{doi:10.1016/j.nima.2011.07.012}}.

\bibitem{bib:cloud}
T.~Tomida, et~al., {Atmospheric monitor for Telescope Array experiment}, EPJ
  Web Conf. 53 (2013) 10003.
\newblock \href {http://dx.doi.org/10.1051/epjconf/20135310003}
  {\path{doi:10.1051/epjconf/20135310003}}.

\bibitem{bib:TA_IMC}
T.~Fujii, et~al., {An event reconstruction method for the Telescope Array
  Fluorescence Detectors}, AIP Conf.Proc. 1367 (2011) 149--152.
\newblock \href {http://dx.doi.org/10.1063/1.3628732}
  {\path{doi:10.1063/1.3628732}}.

\bibitem{bib:gh_function}
T.~K.~Gaisser, A.~M.~Hillas, {Reliability of the method of constant intensity
  cuts for reconstructing the average development of vertical showers},
  International Cosmic Ray Conference, 15th ICRC (Plovdiv) 8 (1977) 353--357.

\bibitem{bib:kakimoto_yield}
F.~Kakimoto, E.~Loh, M.~Nagano, H.~Okuno, M.~Teshima, et~al., {A Measurement of
  the air fluorescence yield}, Nucl.Instrum.Meth. A372 (1996) 527--533.
\newblock \href {http://dx.doi.org/10.1016/0168-9002(95)01423-3}
  {\path{doi:10.1016/0168-9002(95)01423-3}}.

\bibitem{bib:flash}
R.~Abbasi, T.~Abu-Zayyad, K.~Belov, J.~Belz, Z.~Cao, et~al., {Air fluorescence
  measurements in the spectral range 300-420 nm using a 28.5-GeV electron
  beam}, Astropart.Phys. 29 (2008) 77--86.
\newblock \href {http://arxiv.org/abs/0708.3116} {\path{arXiv:0708.3116}},
  \href {http://dx.doi.org/10.1016/j.astropartphys.2007.11.010}
  {\path{doi:10.1016/j.astropartphys.2007.11.010}}.

\bibitem{bib:nkg1}
K.~Kamata, J.~Nishimura, {The Lateral and the Angular Structure Functions of
  Electron Showers}, Prog.Theor.Phys.Suppl. 6 (1958) 93--155.
\newblock \href {http://dx.doi.org/10.1143/PTPS.6.93}
  {\path{doi:10.1143/PTPS.6.93}}.

\bibitem{bib:nkg2}
K.~Greisen, {Cosmic ray showers}, Ann.Rev.Nucl.Part.Sci. 10 (1960) 63--108.
\newblock \href {http://dx.doi.org/10.1146/annurev.ns.10.120160.000431}
  {\path{doi:10.1146/annurev.ns.10.120160.000431}}.

\bibitem{bib:corsika}
D.~Heck, G.~Schatz, T.~Thouw, J.~Knapp, J.~Capdevielle, {CORSIKA: A Monte Carlo
  code to simulate extensive air showers}, Forschungszentrum Karlsruhe Report
  FZKA (1998) 6019.

\bibitem{bib:hires_mia2005}
R.~U. Abbasi, et~al., {A Study of the composition of ultrahigh energy cosmic
  rays using the High Resolution Fly's Eye}, Astrophys. J. 622 (2005) 910--926.
\newblock \href {http://arxiv.org/abs/astro-ph/0407622}
  {\path{arXiv:astro-ph/0407622}}, \href {http://dx.doi.org/10.1086/427931}
  {\path{doi:10.1086/427931}}.

\bibitem{bib:hires_mia}
T.~Abu-Zayyad, et~al., {Measurement of the cosmic ray energy spectrum and
  composition from $10^{17}$ eV to $10^{18.3}$ eV using a hybrid fluorescence
  technique}, Astrophys.J. 557 (2001) 686--699.
\newblock \href {http://arxiv.org/abs/astro-ph/0010652}
  {\path{arXiv:astro-ph/0010652}}, \href {http://dx.doi.org/10.1086/322240}
  {\path{doi:10.1086/322240}}.

\bibitem{bib:hires_gzk}
R.~Abbasi, et~al., {First observation of the Greisen-Zatsepin-Kuzmin
  suppression}, Phys.Rev.Lett. 100 (2008) 101101.
\newblock \href {http://arxiv.org/abs/astro-ph/0703099}
  {\path{arXiv:astro-ph/0703099}}, \href
  {http://dx.doi.org/10.1103/PhysRevLett.100.101101}
  {\path{doi:10.1103/PhysRevLett.100.101101}}.

\bibitem{bib:mass_auger}
A.~Aab, et~al., {Depth of maximum of air-shower profiles at the Pierre Auger
  Observatory. I. Measurements at energies above $10^{17.8}$ eV}, Phys.Rev.
  D90~(12) (2014) 122005.
\newblock \href {http://arxiv.org/abs/1409.4809} {\path{arXiv:1409.4809}},
  \href {http://dx.doi.org/10.1103/PhysRevD.90.122005}
  {\path{doi:10.1103/PhysRevD.90.122005}}.

\bibitem{bib:mass_implication_auger}
A.~Aab, et~al., {Depth of maximum of air-shower profiles at the Pierre Auger
  Observatory. II. Composition implications}, Phys.Rev. D90~(12) (2014) 122006.
\newblock \href {http://arxiv.org/abs/1409.5083} {\path{arXiv:1409.5083}},
  \href {http://dx.doi.org/10.1103/PhysRevD.90.122006}
  {\path{doi:10.1103/PhysRevD.90.122006}}.

\bibitem{bib:auger_mass_icrc2015}
A.~Porcelli, et~al., {Measurements of $X_{\max}$ above 10$^{17}$ eV with the
  fluorescence detector of the Pierre Auger Observatory}, Proc. of the 34th
  International Cosmic Ray Conference, Hague, Netherland, PoS (ICRC 2015) 420.
\newblock \href {http://arxiv.org/abs/1509.03732} {\path{arXiv:1509.03732}}.

\bibitem{bib:composition_wg}
E.~Barcikowski, et~al., {Mass Composition Working Group Report at UHECR-2012},
  EPJ Web Conf. 53 (2013) 01006.
\newblock \href {http://arxiv.org/abs/1306.4430} {\path{arXiv:1306.4430}},
  \href {http://dx.doi.org/10.1051/epjconf/20135301006}
  {\path{doi:10.1051/epjconf/20135301006}}.

\bibitem{bib:composition_wg2014}
R.~Abbasi, et~al., {Report of the Working Group on the Composition of Ultra
  High Energy Cosmic Rays}, Proc. of 2014 Conference on Ultrahigh Energy Cosmic
  Rays, USA, Utah (2014)\href {http://arxiv.org/abs/1503.07540}
  {\path{arXiv:1503.07540}}.

\bibitem{bib:kascade_light}
W.~D. Apel, et~al., {Ankle-like Feature in the Energy Spectrum of Light
  Elements of Cosmic Rays Observed with KASCADE-Grande}, Phys. Rev. D87 (2013)
  081101.
\newblock \href {http://arxiv.org/abs/1304.7114} {\path{arXiv:1304.7114}},
  \href {http://dx.doi.org/10.1103/PhysRevD.87.081101}
  {\path{doi:10.1103/PhysRevD.87.081101}}.

\bibitem{bib:kascade_icrc2015}
S.~Schoo, et~al., {The energy spectrum of cosmic rays in the range from
  10$^{14}$ to 10$^{18}$ eV}, Proc. of the 34th International Cosmic Ray
  Conference, Hague, Netherland, PoS (ICRC 2015) 263.

\bibitem{bib:sean_mc}
S.~R. Stratton, D.~R. Bergman, T.~A. Stroman, L.~M. Scott, S.~Ogio, T.~Fujii,
  Y.~Tsunesada, {Using the Monte Carlo Technique in the Observation of
  Fluorescence from UHECRs}, Proc of the 32nd International Cosmic Ray
  Conference, Beijing, China, 2 (2011) 262.
\newblock \href {http://dx.doi.org/10.7529/ICRC2011/V02/1299}
  {\path{doi:10.7529/ICRC2011/V02/1299}}.

\bibitem{bib:icetop-73}
M.~G. Aartsen, et~al., {Measurement of the cosmic ray energy spectrum with
  IceTop-73}, Phys. Rev. D88~(4) (2013) 042004.
\newblock \href {http://arxiv.org/abs/1307.3795} {\path{arXiv:1307.3795}},
  \href {http://dx.doi.org/10.1103/PhysRevD.88.042004}
  {\path{doi:10.1103/PhysRevD.88.042004}}.

\bibitem{bib:kascade_spectrum}
W.~D. Apel, et~al., {The spectrum of high-energy cosmic rays measured with
  KASCADE-Grande}, Astropart. Phys. 36 (2012) 183--194.
\newblock \href {http://dx.doi.org/10.1016/j.astropartphys.2012.05.023}
  {\path{doi:10.1016/j.astropartphys.2012.05.023}}.

\bibitem{bib:AugerSDSpectrum}
I.~Valino, et~al., {The flux of ultra-high energy cosmic rays after ten years
  of operation of the Pierre Auger Observatory}, Proc. of the 34th
  International Cosmic Ray Conference, Hague, Netherland, PoS (ICRC 2015) 271.
\newblock \href {http://arxiv.org/abs/1509.03732} {\path{arXiv:1509.03732}}.

\bibitem{bib:tamd_spectrum}
T.~Abu-Zayyad, et~al., {The Energy Spectrum of Telescope Array's Middle Drum
  Detector and the Direct Comparison to the High Resolution Fly's Eye
  Experiment}, Astropart.Phys. 39-40 (2012) 109--119.
\newblock \href {http://arxiv.org/abs/1202.5141} {\path{arXiv:1202.5141}},
  \href {http://dx.doi.org/10.1016/j.astropartphys.2012.05.012}
  {\path{doi:10.1016/j.astropartphys.2012.05.012}}.

\bibitem{bib:airfly1}
M.~Ave, et~al., {Precise measurement of the absolute fluorescence yield of the
  337 nm band in atmospheric gases}, Astropart.Phys. 42 (2013) 90--102.
\newblock \href {http://arxiv.org/abs/1210.6734} {\path{arXiv:1210.6734}},
  \href {http://dx.doi.org/10.1016/j.astropartphys.2012.12.006}
  {\path{doi:10.1016/j.astropartphys.2012.12.006}}.

\bibitem{bib:airfly2}
M.~Ave, et~al., {Measurement of the pressure dependence of air fluorescence
  emission induced by electrons}, Astropart.Phys. 28 (2007) 41--57.
\newblock \href {http://arxiv.org/abs/astro-ph/0703132}
  {\path{arXiv:astro-ph/0703132}}, \href
  {http://dx.doi.org/10.1016/j.astropartphys.2007.04.006}
  {\path{doi:10.1016/j.astropartphys.2007.04.006}}.

\end{thebibliography}

\section*{Appendix: Spectrum data}
The energy spectrum measured by the BRM and LR stations is listed in Table~\ref{tbl:spectrum_data}. The number of events in each energy bin is also indicated.
The total systematic uncertainties are calculated by uncertainties of the mass composition and the energy scale in quadrature.
\begin{table}
  \centering
  \begin{tabular}{lrr} \hline
$\frac{\rm{Energy \, bin}}{\rm{(eV)}}$  & $N$  & $\frac{J(E) E^3\rm{+\sigma_{stat.}-\sigma_{stat.}+\sigma_{syst.}-\sigma_{syst.}}}{\rm{ \times 10^{24} (eV^{2} m^{-2} sr^{-1} s^{-1})}}$   \\ \hline
17.2-17.3  &  858  &  2.95 + 0.30 - 0.30 + 2.04 - 1.13 \\
17.3-17.4  &  1444  &  2.85 + 0.20 - 0.20 + 1.62 - 1.07 \\
17.4-17.5  &  2252  &  2.96 + 0.14 - 0.14 + 1.44 - 1.09 \\
17.5-17.6  &  2820  &  2.79 + 0.10 - 0.10 + 1.22 - 1.01 \\
17.6-17.7  &  3279  &  2.73 + 0.08 - 0.08 + 1.11 - 0.98 \\
17.7-17.8  &  3216  &  2.47 + 0.06 - 0.06 + 0.95 - 0.88 \\
17.8-17.9  &  3100  &  2.38 + 0.06 - 0.06 + 0.89 - 0.84 \\
17.9-18.0  &  2716  &  2.25 + 0.05 - 0.05 + 0.82 - 0.79 \\
18.0-18.1  &  2298  &  2.17 + 0.06 - 0.06 + 0.78 - 0.76 \\
18.1-18.2  &  1724  &  1.94 + 0.07 - 0.07 + 0.69 - 0.68 \\
18.2-18.3  &  1311  &  1.84 + 0.07 - 0.07 + 0.65 - 0.64 \\
18.3-18.4  &  965  &  1.74 + 0.07 - 0.07 + 0.62 - 0.61 \\
18.4-18.5  &  737  &  1.77 + 0.08 - 0.07 + 0.62 - 0.62 \\
18.5-18.6  &  448  &  1.50 + 0.08 - 0.08 + 0.53 - 0.52 \\
18.6-18.7  &  332  &  1.63 + 0.10 - 0.09 + 0.57 - 0.57 \\
18.7-18.8  &  204  &  1.49 + 0.12 - 0.11 + 0.52 - 0.52 \\
18.8-18.9  &  177  &  1.96 + 0.16 - 0.15 + 0.69 - 0.69 \\
18.9-19.0  &  109  &  1.85 + 0.20 - 0.18 + 0.65 - 0.65 \\
19.0-19.1  &  86  &  2.25 + 0.27 - 0.25 + 0.79 - 0.79 \\
19.1-19.2  &  70  &  2.84 + 0.39 - 0.35 + 1.00 - 1.00 \\
19.2-19.3  &  32  &  2.03 + 0.43 - 0.36 + 0.71 - 0.71 \\
19.3-19.4  &  37  &  3.68 + 0.71 - 0.61 + 1.29 - 1.29 \\
19.4-19.5  &  25  &  3.90 + 0.95 - 0.78 + 1.37 - 1.37 \\
19.5-19.6  &  8  &  1.97 + 0.97 - 0.68 + 0.69 - 0.69 \\
19.6-19.7  &  7  &  2.71 + 1.46 - 1.00 + 0.95 - 0.95 \\
19.7-19.8  &  7  &  4.28 + 2.31 - 1.58 + 1.50 - 1.50 \\
19.8-19.9  &  6  &  5.80 + 3.46 - 2.31 + 2.03 - 2.03 \\
19.9-20.0  &  1  &  1.53 + 3.54 - 1.27 + 0.54 - 0.54 \\
20.0-20.1  &  1  &  2.42 + 5.60 - 2.00 + 0.85 - 0.85 \\ \hline 
 \end{tabular}
\caption{Energy bin, number of events in the bin, $N$, and energy spectrum data.}
\label{tbl:spectrum_data}
\end{table}

\end{document}